\documentclass[aps,pre,twocolumn,superscriptaddress,showpacs]{revtex4}
\usepackage{graphicx}
\usepackage{amsmath}
\usepackage{subfigure}
\usepackage{epsfig} 
\usepackage{epstopdf}
\usepackage{float} 
\usepackage{tikz}
\usepackage[percent]{overpic}
\usepackage{comment}
\usepackage{pstricks}
\usepackage{placeins}

\begin{document}

\title{Signatures of Chaos in Time Series Generated by Many-Spin Systems at High Temperatures }



\author{Tarek A. Elsayed}
\email{T.Elsayed@thphys.uni-heidelberg.de}
\address{Institute for Theoretical Physics, University of Heidelberg, Philosophenweg 19, 69120 Heidelberg, Germany}
\author{Benjamin Hess}
\address{Institute for Theoretical Physics, University of Heidelberg, Philosophenweg 19, 69120 Heidelberg, Germany}
\author{Boris V. Fine}
\email{B.Fine@thphys.uni-heidelberg.de}
\address{Institute for Theoretical Physics, University of Heidelberg, Philosophenweg 19, 69120 Heidelberg, Germany}
\address{Department of Physics, School of Science and Technology, Nazarbayev University,
53 Kabanbai Batyr Ave., Astana 010000, Kazakhstan}


\date{\today}

\begin{abstract}
Extracting reliable indicators of chaos from a single experimental time series is a  challenging task, in particular, for systems with many degrees of freedom. The techniques available for this purpose often require unachievably long time series. In this paper, we explore a new method of discriminating chaotic from multi-periodic integrable motion in many-particle systems. The applicability of this method is supported by our numerical simulations of the dynamics of classical spin lattices at high temperatures. 
We compared chaotic and nonchaotic regimes of these lattices and investigated the transition between the two. 
The method is based on analyzing higher-order time derivatives of the time series of a macroscopic observable --- the total magnetization of the spin lattice. We exploit the fact that power spectra of the magnetization time series generated by chaotic spin lattices exhibit exponential high-frequency tails, while, for the integrable spin lattices, the power spectra are terminated in a non-exponential way.
We have also demonstrated the applicability limits of the above method by investigating the high-frequency tails of the power spectra generated by quantum spin lattices and by the classical Toda lattice. 
\end{abstract}

\maketitle

\section{Introduction}

The problem of detecting deterministic chaos in an experimental time series is of fundamental interest for the foundations of statistical physics \cite{gaspard-93, cecconi}. It also arises in other contexts \cite{kantz,Argyris-94}, e.g., in biomedical applications \cite{heart}. In the context of statistical physics, the presence of many degrees of freedom poses a formidable challenge to the investigations of chaos in various systems such as, for example, lattices of nonlinearly coupled oscillators\cite{Matthews-91,Berman-05}. 
An often encountred difficulty for the time series analysis  is to determine whether a given many-particle system is chaotic by accessing the time evolution of only one degree of freedom. In such a case, the primary indicator of chaos, namely, the Lyapunov instability in the many-dimensional phase space cannot be investigated directly. 
A notable example in this regard was an attempt of Ref. \cite{gaspard-98} to identify microscopic chaos in the measured trace of a Brownian particle. The approach of Ref. \cite{gaspard-98} was to analyze the rate of information entropy (IE) production by this trace. The limiting value of this rate in the chaotic systems is known to be equal to the sum of the positive Lyapunov exponents.  The results of Ref. \cite{gaspard-98}  were consistent with the possible presence of microscopic chaos, but, at the same time, were criticized as leaving open the possibility that the same signatures might be produced by nonchaotic systems \cite{dettmann,schreiber-99}.

  Detection of microscopic chaos means that (i) it should be discriminated from  a stochastic noise process characterized by the infinite limiting rate of IE-production and  (ii) it should be discriminated from a multi-periodic integrable motion characterized by the zero rate of IE-production. The difficulty here is that extracting the true limiting behavior of the rate of IE-production in many-particle systems typically requires unachievably long time series. In the present paper, we focus primarily on issue (ii) above for time series of magnetization produced by a macroscopic system of classical spins.  We also investigate the classical Toda lattice and quantum lattices of spins 1/2.
  
    We propose a new method to discriminate chaos from a multi-periodic nonchaotic motion in a short very accurately measured time series. We consider a  many-dimensional Hamiltonian system, where both chaotic and nonchaotic motions are smooth in time, and the nonchaotic motion is characterized by a sufficiently large number of frequencies that cannot be resolved by the Fourier transform of the time series. The method is based on analyzing the higher-order time derivatives of the time series. It exploits the fact, which we demonstrate numerically, that the power spectra of time series generated by chaotic {\it many-spin} Hamiltonian systems have exponential high-frequency tails. To the best of our knowledge, the existence of such tails in generic chaotic systems has never been proven rigorously, but otherwise reported in the studies of non-Hamiltonian or low-dimensional chaotic systems \cite{farmer-82,frisch-81,sigeti-pre-95}.  In contrast, the power spectra of many integrable systems, including integrable spin systems, are  normally  terminated faster than an exponential function, and their high-frequency behavior is non-universal. Taking time derivatives of a time series progressively suppresses the low frequency parts of the power spectra and amplifies the high-frequency part, thereby producing a number of clearly detectable signatures of the exponential tail of the power spectrum.

 Particularly intriguing is the possibility to apply this method to the time series obtained by monitoring large quantum systems. Should a quantum time series produce the same signatures of chaos as expected for classical systems, such a result would shed a new light on the notion of quantum chaos.
It would also be consistent with the previous studies linking the generic functional form of the long-time relaxation in classical and quantum spin systems to  microscopic chaos \cite{fine-03,fine-04,fine-08,sorte}.

Much of the content of this paper concerns with the applicability limits of the above method of detecting chaos.  Postponing the discussion of this issue until Section~\ref{discussion}, we just mention here that, since there is no known quantitative connection between the high-frequency tails of the power spectra and the values of the Lyapunov exponents,  using the shape of these tails as a criterion of chaos has limitations. Rigourously speaking, the exponential shape of the high-frequency tails of the power spectra is neither necessary nor sufficient condition of chaos. In this paper, the applicability of the above criterion can be considered as reliably established only for 
classical spin lattices with nearest neighbor interaction at high temperatures. However, we expect that this exponential shape can be used as an empirical criterion of chaos also for a broader class of many-particle systems, whose dynamics is sufficiently similar to that of the classical spin lattices.

The rest of the paper is organized as follows. In Section~\ref{lattices} , we define an integrable and a nonintegrable models of classical spin clusters. In Section~\ref{numerical}, we  demonstrate the chaotic character of the nonintegrable cluster by calculating its largest Lyapunov exponent.  In Section~\ref{rate}, we consider rather long time series for the two clusters and demonstrate that one cannot discriminate between the chaotic and nonchaotic time series on the basis of the numerically accessible rate of IE-production. In Section~\ref{long}, we proceed with calculating the higher-order time derivatives for the two time series and show that their seventh derivatives already look qualitatively different so that one can discriminate the chaotic from the nonchaotic time series simply by visual inspection.  We further introduce several quantitative criteria characterizing this difference.  In Sections~\ref{Noisy} and \ref{Short}, we also demonstrate the utility of our method by applying it to noisy time series and to a very short time series.   In Section~\ref{transition}, we illustrate the qualitative change of the exponential tail of the power spectrum during the transition to integrability. 
Finally, we investigate the applicability of the above method beyond classical spin lattices, and report both supporting and contradictig evidence obtained by calculating the high-frequency tales of the power spectra generated by the completely integrable Toda lattice (Section~\ref{Toda}) and by  quantum spin 1/2 lattices (Section~\ref{quantum}). The overall discussion of the applicability of the method is given in Section~\ref{discussion}. 
 
\section{Lattices of classical spins}
\label{lattices}



Lattices of classical spins  are generically chaotic  but the interaction constants can also be chosen such that the spin dynamics becomes nonchaotic \cite{fine-12, fine2013}. 

We initially consider two $6 \times 6 \times 6$ spin clusters characterized by the nearest-neighbor (NN) interaction Hamiltonian
\begin{equation}\label {hamiltonian}
H=\sum_{i<j}^{\hbox{\scriptsize NN}} J_x S_{ix} S_{jx}+J_y S_{iy} S_{jy}+J_z S_{iz} S_{jz},
\end{equation}
where $(S_{ix},S_{iy},S_{iz})$ are the components of the $i^\text{th}$ classical spin normalized by the condition $S_{ix}^2 + S_{iy}^2 + S_{iz}^2 = 1$, and 
$J_x, J_y$ and $J_z$ are the coupling constants. For the integrable cluster, we select the Ising Hamiltonian characterized by \( J_x = J_y = 0\) and \( J_z=1\), while, for the chaotic cluster, we select \( J_x =-0.65, J_y = -0.3 , J_z=0.7\).  The characteristic time scale of the dynamics of both clusters is made equal since, in both cases, \( J_x^2+J_y^2+J_z^2 =1 \) .

 In the Ising case, which is  the only integrable case of cubic classical spin clusters with  nearest neighbour interactions \cite{fine-12}, the motion is integrable because the $z$-component of each spin is a constant of motion, while the $x$- and the $y$-components simply precess in the local fields created by the frozen $z$-components of the neighbors. Therefore, the nonchaotic time series to be computed below is characterized by the superposition of 216 different frequencies. 
 
Despite the rather finite size of the above clusters, we expect the numerical results reported in the rest of the paper to represent lattices of infinite size. The reason is that the timescales relevant to the reported phenomenology, namely, the inverse largest Lyapunov exponent and the inverse characteristic frequency scales of the power spectra are intensive quantities: they stop changing rather quickly as the size of the system increases. For the power spectra of nonconserved projections of the total magnetization, this was demostrated numerically in Ref.\cite{fine-03}. The intensive character of the largest Lyapunov exponents for classical spin lattices with nearest neighbor interactions is rigorously justified in Ref.\cite{fine2013}.
 
 
\section{Numerical procedures}
\label{numerical}

The equations of motion for the Hamiltonian (\ref{hamiltonian}) were solved numerically using the discretization routine of Ref. \cite{fine-03}. This routine conserves the energy of the system exactly. The typical discretization time step $\Delta t$ was 0.01. The discretization errors were controlled by examining the effects of the change of $\Delta t$ on the Lyapunov exponents and on the power spectra of the time series.  The initial orientations of spins were chosen randomly. Therefore, the energy of the system was close to zero.

We computed the largest Lyapunov exponent for both clusters using the method of Ref. \cite{benettin}. The method consists of choosing a small initial  distance $d_0$ between two phase space trajectories, letting them diverge during time $\tau$ and then resetting this distance back to $d_0$ along the displacement direction just before the reset, and so on, repeating the above manipulation many times. If the system is chaotic, the spread of the two trajectories is eventually controlled by the largest Lyapunov exponent, which can be calculated as the limiting value of the expression  
\begin{equation}
\lambda _{\hbox{\scriptsize max}}=\frac{1}{k \tau}\sum_{j=1}^{k} \hbox{log}\left ( \frac{d_{j}}{d_{0}} \right ),
\label{lambdamax}
\end{equation}
 where $j$ is the reset index, $k$ is the total number of resets, and $d_j$ is the distance between the two trajectories just before the $j^\text{th}$ reset. 
 
 For a fixed value of $\tau$, integrable systems can also produce small but finite value of $\lambda _{\hbox{\scriptsize max}}$ because of the polynomial spread of the trajectories. However, for the polynomial spread,  $\lambda _{\hbox{\scriptsize max}} \sim 1/\tau$ (up to a logarithmic prefactor), while for the exponential spread the value of $\lambda _{\hbox{\scriptsize max}}$ should not depend on $\tau$. 
 
\begin{figure} \setlength{\unitlength}{0.1cm}

\begin{picture}(90 , 27 ) 
{
\put(-0.64, 25){{ (a)}}
\put(0, 0){  \epsfig{file=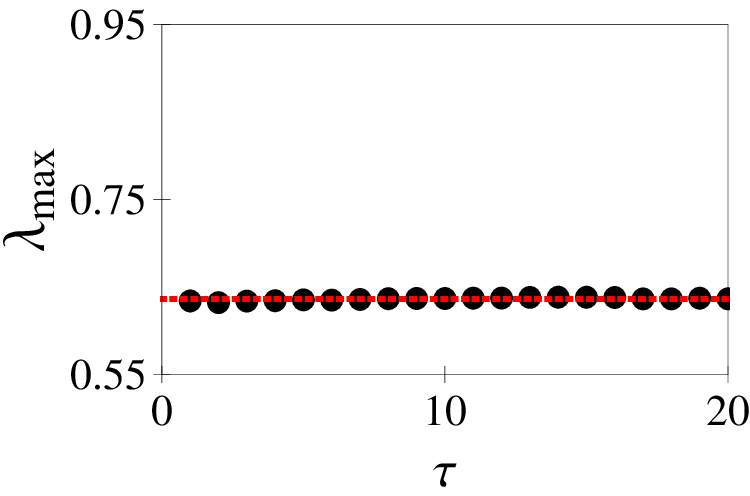,width=4.2cm } }
\put(43, 0){ \epsfig{file=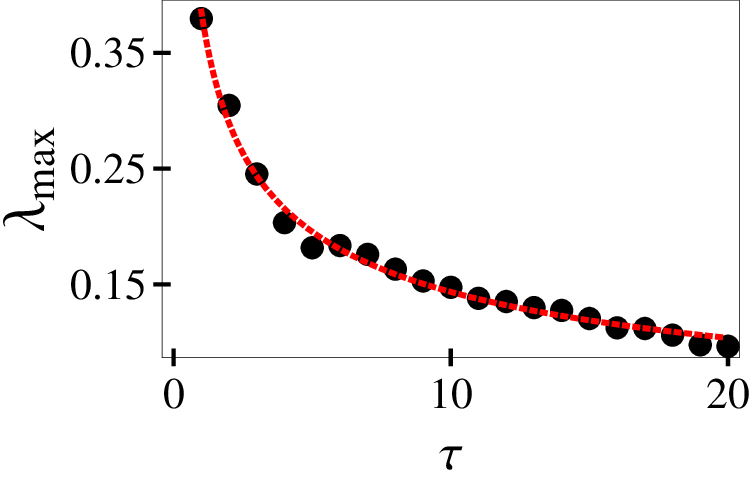,width=4.2cm  } }
\put(42.1, 25){{ (b)}}
\put(21.5, 13){  \epsfig{file=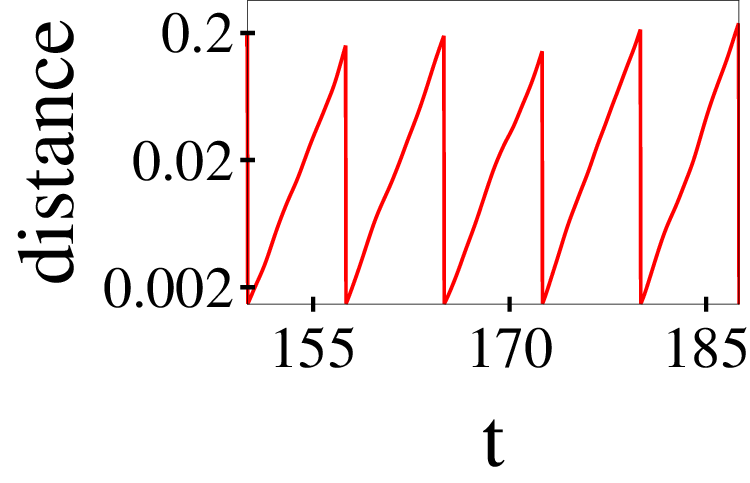,width=1.8cm } }
\put(63.5, 12){ \epsfig{file=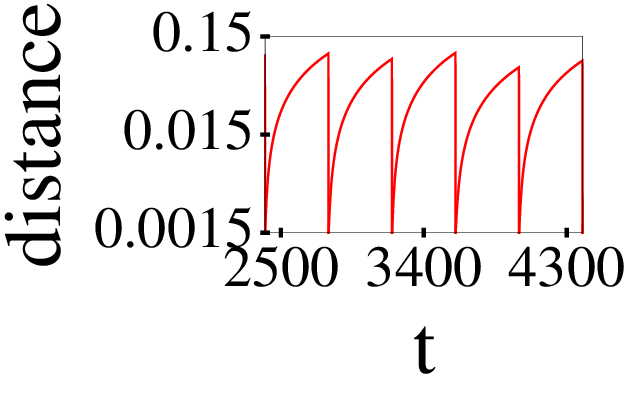,width=2cm  } }
}
\end{picture}

\caption{\label{lyp} (Color online)  Numerical values of the maximum Lyapunov exponent \( \lambda _\text{max} \) for the chaotic (a) and the integrable (b) systems as a function of the reset time $\tau$ (dots). The lines represent the fits by a constant in (a) and by function \( \frac{\log(1+\tau ^{\beta })}{\tau } \) in (b). The insets illustrate the growth of the distance between a pair of trajectories during 5 reset intervals. }
\end{figure}

 The dependences of $\lambda _{\hbox{\scriptsize max}}$ on $\tau$ for the two clusters are shown in Fig.~\ref{lyp}. For the Ising cluster, $\lambda _{\hbox{\scriptsize max}}$ approaches zero approximately as $1/\tau$ as expected for an integrable system. The chaotic cluster exhibits $\tau$-independent value $\lambda _{\hbox{\scriptsize max}} = 0.63$. The insets of Figs.\ref{lyp}(a) and (b) show that distance growth between two resets is exponential in the chaotic case, and slower than exponential in the integrable case.

\section{Rate of information entropy production}
\label{rate}

In this and the next section, we try discriminate two time series of the total magnetization, $M_x$, for the chaotic and nonchaotic clusters introduced in section II. The length of each time series is $T=1000$. The two time series look very similar as  illustrated in Figs.~\ref{main} (a) and (b).

\begin{figure} 
\includegraphics[scale=0.5]{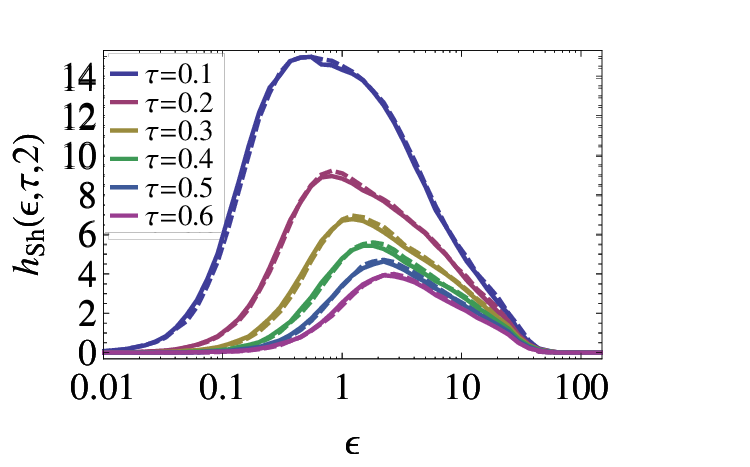}
\caption{ \label{entropy} (Color online) Comparison between IE rates $\mathit{h}_{Sh}(\epsilon, \tau,N)$  produced by the chaotic system (solid lines) and by the integrable system (dashed lines) for $N=2$ and different values of \(\tau\).}
\end{figure}

In the present section, we investigate the rates of the \( (\varepsilon ,\tau)  \)-information-entropy production \cite{gaspard-93,cencini,abel2000,Abel-00} for the two time series. For this purpose, we ``coarse-grained" the time and the magnetization axes in steps of \(\tau\) and \( \varepsilon\), respectively, to obtain a newly discretized version of the time series (a stream of symbols). The \( (\varepsilon,\tau) \) Shannon information entropy for patterns of length $N$ is given by
$
\mathit{H_{Sh}}(\varepsilon,\tau,N)= - \sum \mathit{P_i} \ \hbox{log} \ P_i ,
$
where $i$ is the index of distinct patterns, and $P_i$ is the pattern probability. The Shannon \( (\varepsilon,\tau) \)-entropy per unit time is defined as 
$ \lim_{N \to \infty } \mathit{h}_{Sh}(\epsilon, \tau, N)$, where 
$
\mathit{h}_{Sh}(\epsilon, \tau,N)=\frac{1}{\tau}\  [\mathit{H}_{Sh}(\varepsilon,\tau,N+1)-\mathit{H}_{Sh}(\varepsilon,\tau,N)]
$.
For a chaotic system, $h_{Sh}(\epsilon, \tau, N)$ is expected to approach the constant value equal to the sum of the positive Lyapunov exponents in the limit $T \to \infty$, $N \to \infty$, $\varepsilon \to 0$ and $\tau \to 0$. However, for many-dimensional systems, the above limit is, typically, impossible to reach in practice, because, as $N$ increases or $\varepsilon$ decreases, any finite time series quickly becomes too short to fairly represent the statistics of all possible patterns of length $N$. When this happens, each pattern, that occurs, occurs only once, and, as a result,
$h_{Sh}(\epsilon, \tau, N) \to 0$. We, nevertheless, calculated $h_{Sh}(\epsilon, \tau, 2)$  in order to check if there is any  indication of chaos before the effect of the finite length of the time series sets in.
The results presented in Fig.~\ref{entropy} are nearly identical for both chaotic and nonchaotic time series. 

\begin{figure}
\includegraphics[scale=0.6]{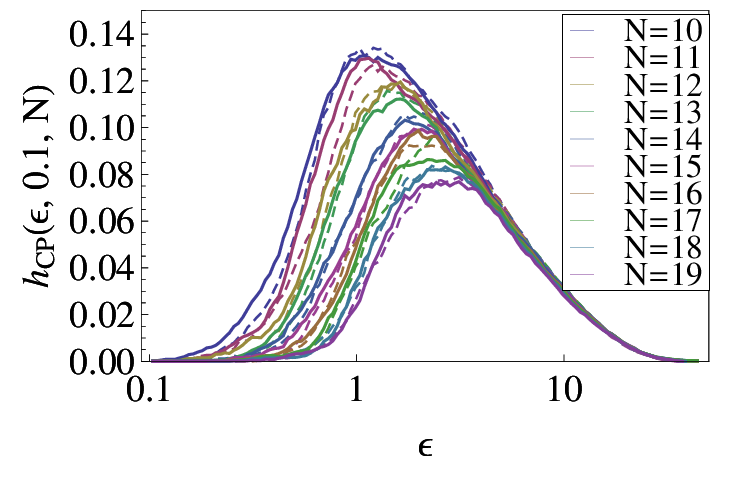}
\caption
{ (Color online) Comparison between $\mathit{h}_{CP}(\epsilon, \tau,N)$ with $\tau=0.1$ and various values of $N$ for the chaotic time series (solid lines) and the integrable time series (dashed lines). Curves reaching higher maximum values of $\mathit{h}_{CP}$ correspond to the smaller values of $N$.} 
\label{CPentropy}
\end{figure}

Similar results were also obtained for the Cohen-Procaccia (CP) entropy \cite{CP-pra}. 
The magnetization axis was not discretized for the calculation of the CP 
$(\varepsilon ,\tau)$-entropy. Instead, the distance between two patterns of length $N$ was  computed as  the maximum of the differences between each pair of time-discretized points. To calculate the CP $(\varepsilon ,\tau)$-entropy, \( \mathit{H_{CP}}(\varepsilon,\tau,N) \), we selected a group of $R=100$ reference patterns of length $N$ randomly, and then obtained the probability of each of them by counting the number of all other patterns occurring in the time series which are within distance \(  \varepsilon \) from that reference pattern. The CP $(\varepsilon ,\tau)$-entropy was computed using the formula 
\begin{equation}
\mathit{H_{CP}}(\varepsilon,\tau,N)=-\frac{1}{R}\sum_{\left \{ R \right \}} \hbox{log} P_i.
\end{equation}
The limiting rate of the CP $(\varepsilon ,\tau)$-entropy production is defined as $\lim_{N \to \infty } \mathit{h}_{CP}(\epsilon, \tau,N)$, where
\begin{equation}\label{CPrate}
\mathit{h}_{CP}(\epsilon, \tau,N)=\frac{1}{\tau}\ [\mathit{H}(\varepsilon,\tau,N+1)-\mathit{H}(\varepsilon,\tau,N)].
\end{equation}
As in the case of Shannon entropy, we were not able to approach the true limit $\lim_{N \to \infty } \mathit{h}_{CP}(\epsilon, \tau,N)$ numerically. Instead, we obtained  $\mathit{h}_{CP}(\epsilon, \tau,N)$ for the values of $N$ up $N=19$.  In Fig. \ref{CPentropy}, we show the CP $(\varepsilon,\tau, N)$-entropy plots corresponding to the same data set as those analysed in Fig.~\ref{entropy}. We used  \( \tau = 0.1 \) (10 discretization time steps) for one pattern element and varied $N$ from 10 to 19 in different plots.  It is evident  from Fig. \ref{CPentropy} that one cannot distinguish integrable from chaotic dynamics on the basis of  numerically accessible behavior of $\mathit{h}_{CP}(\epsilon, \tau,N)$. 

\section{Higher-order time derivatives of the magnetization time series }
\label{higher}

\subsection{Long time series without noise}
\label{long}

\begin{figure} \setlength{\unitlength}{0.1cm}

\begin{picture}(90 , 86 ) 
{ 
\put(-0.5, 78){{ (a)}}
\put(42.5, 78){{ (b)}}
\put(0, 54){  \epsfig{file=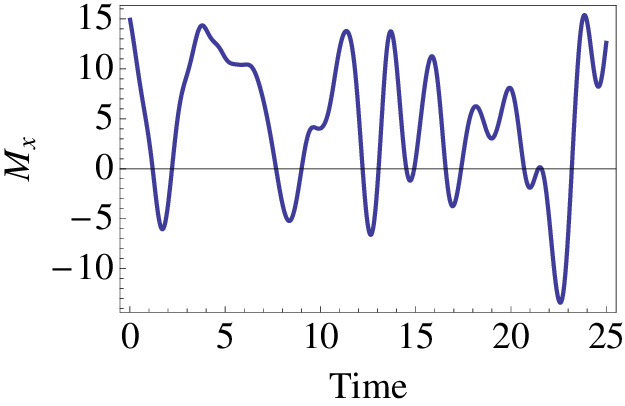,width=4.2cm } }
\put(43, 54){ \epsfig{file=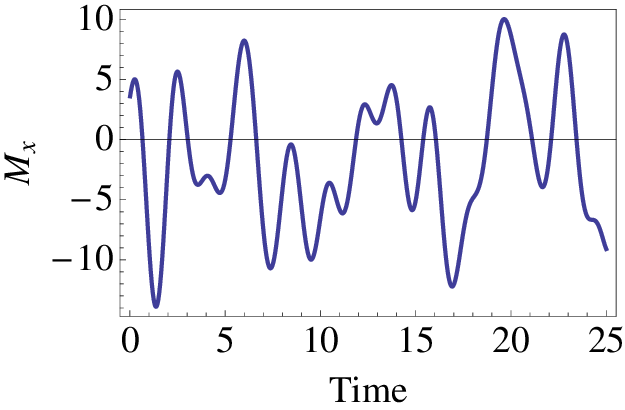,width=4.2cm  } }
\put(0, 27){  \epsfig{file=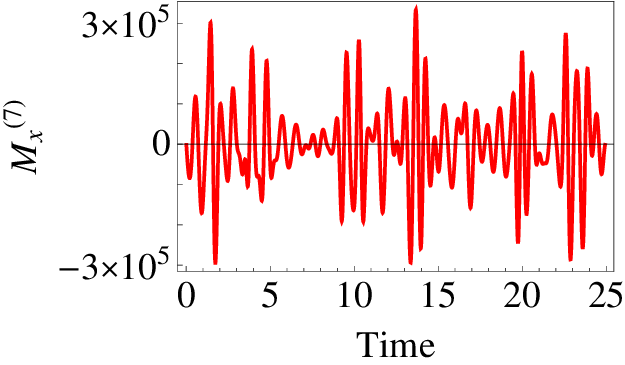,width=4.2cm } }
\put(43, 27){ \epsfig{file=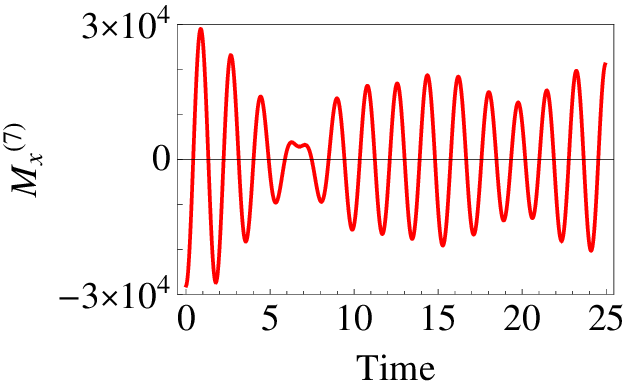,width=4.2cm  } }
\put(-0.5, 52){{ (c)}}
\put(42.5, 52){{ (d)}}
\put(0, -2){  \epsfig{file=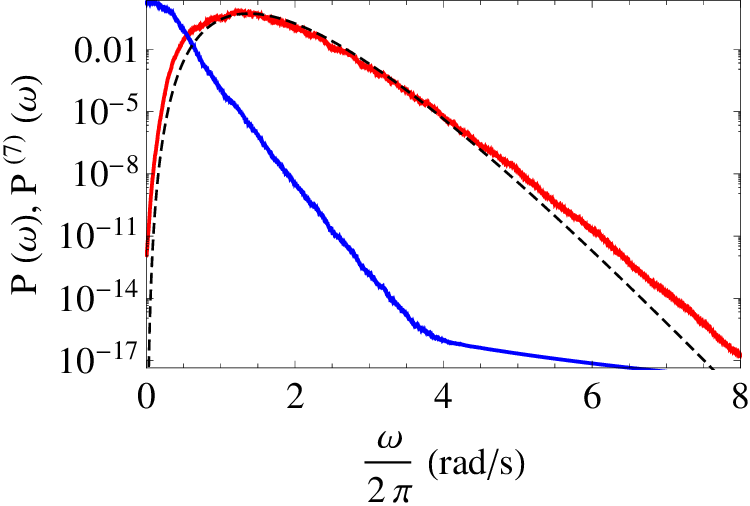,width=4.2cm } }
\put(43, -2){ \epsfig{file=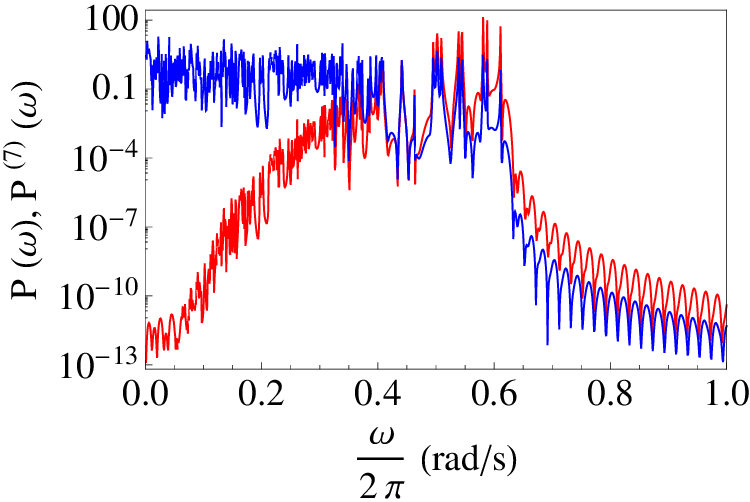,width=4.2cm  } }
\put(0, 24){{ (e)}}
\put(43, 24){{ (f)}}
\put(20,85){Chaotic}
\put(62,85){Integrable}
}
\end{picture} 
\caption{(Color online) Time series of $M_x(t)$ for  $6\times 6 \times 6$ spin clusters. Figures in the left column represent the chaotic cluster: (a) fragment of the time series $M_x(t)$; (c) the corresponding 7th time derivative $M_x^{(7)}(t)$; and (e) the power spectra $P(\omega)$ [blue] and $P^{(7)}(\omega)$[red]. Figures (b,d,f)  in the right column represent the same quantities for the integrable Ising cluster. The black dashed line in (e) depicts fit to $P^{(7)}(\omega)$ of the form  \(P_0 \; \omega^{14} e^{-\gamma \omega } \) where $P_0$ and  $\gamma$ are fitting parameters. The power spectra were obtained as described in Appendix B.  In (e), they  were also smoothed out.  The tails of the power spectra at $\omega/2\pi > 4$ in (e) and $\omega/2\pi > 0.6$ in (f)  are affected by the spectral leakage from lower frequencies due to the finite length of the time series. 
} 
\label{main} 
\end{figure}
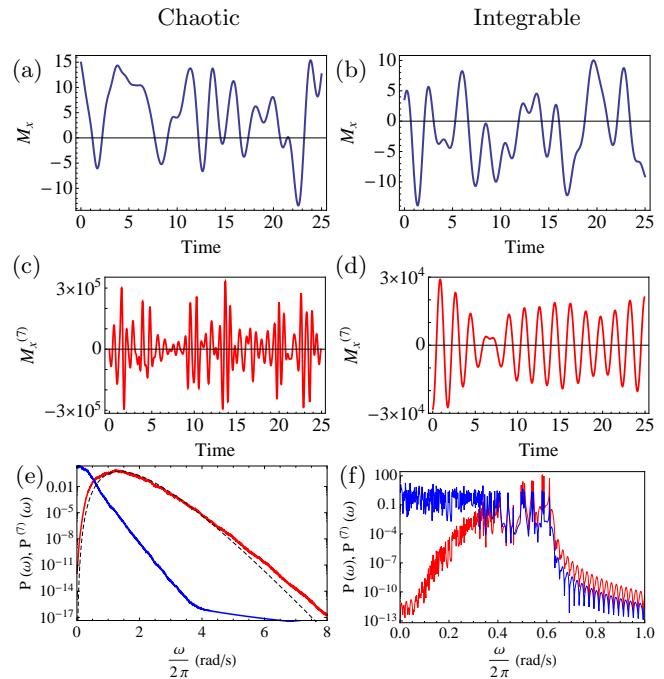

We demonstrate in this section that one can easily distinguish chaotic from  nonchaotic  time series of $M_x(t)$ by looking at its derivatives of the $n$-th order, which we denote as $M_x^{(n)}(t)$. In
Figs.~\ref{main}(c) and (d), we exemplify this statement by presenting the time evolution of the $7^\text{th}$ time derivative $M_x^{(7)}(t)$ for the seemingly indistinguishable time series appearing in Figs.~\ref{main}(a) and (b). (The plots for the lower-order derivatives are given in  Appendix A.) For the chaotic time series,  $M_x^{(7)}(t)$ fluctuates noticeably faster than $M_x(t)$ and has a rather random appearance, while, for the nonchaotic time series, $M_x^{(7)}(t)$ has the appearance of a slowly modulated periodic signal with period of the order of the characteristic time of $M_x(t)$.

The clear difference between Figs.~\ref{main}(c) and (d) can be understood from the fact that the power spectrum of the original time series, $P(\omega)$, and the power spectrum of the $n$th time derivative $P^{(n)}(\omega)$ are related as $P^{(n)}(\omega) = \omega^{2n} P(\omega)$.  Both $P(\omega)$ and $P^{(7)}(\omega)$ for each time series are presented in Figs.~\ref{main}(e) and (f). In the nonintegrable case, $P(\omega)$  has an exponential tail of the form
\begin{equation}
P(\omega)=P_0 e^{-\gamma |\omega|},
\label{p_of_omega}
\end{equation}
 where $\gamma$ and $P_0$ are constants. As a result, for sufficiently large $n$, 
\begin{equation}
P^{(n)}(\omega) \cong  P_0\omega^{2n} e^{-\gamma |\omega|}.
\label{criterion}
\end{equation}
We propose to use this dependence as a criterion of chaos in classical spin systems. Figure~\ref{main}(e) includes the fit to $P^{(7)}(\omega)$ of the form $P_0\ \omega^{14} e^{-\gamma |\omega|}$. The important aspect of this dependence is not that it becomes exponential at sufficiently large frequencies, but that, before it becomes exponential, it has the universally shaped broad maximum shifting with $n$ to increasingly high frequencies. In contrast, the power spectrum for the integrable cluster is terminated sharply at a certain maximum frequency $\omega_{max}$. As a result, the shape of $\omega^{2n} P(\omega)$ for sufficiently large $n$ becomes narrowly peaked around $\omega_{max}$. Therefore, $\omega_{max}$ becomes the carrier frequency for the modulations observed in Fig.~\ref{main}(d), while the inverse width of this peak characterizes the modulation time scale.

\begin{figure} \setlength{\unitlength}{0.1cm}

\begin{picture}(90 , 27 ) 
{

\put(0, 0){  \epsfig{file=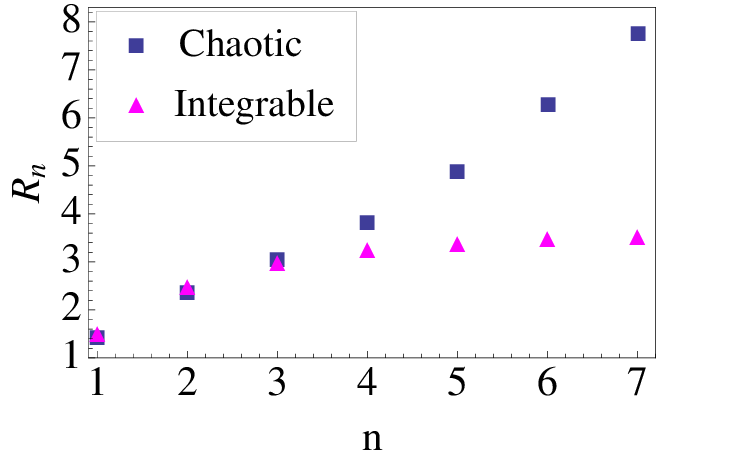,width=4.4cm } }
\put(43, 0){ \epsfig{file=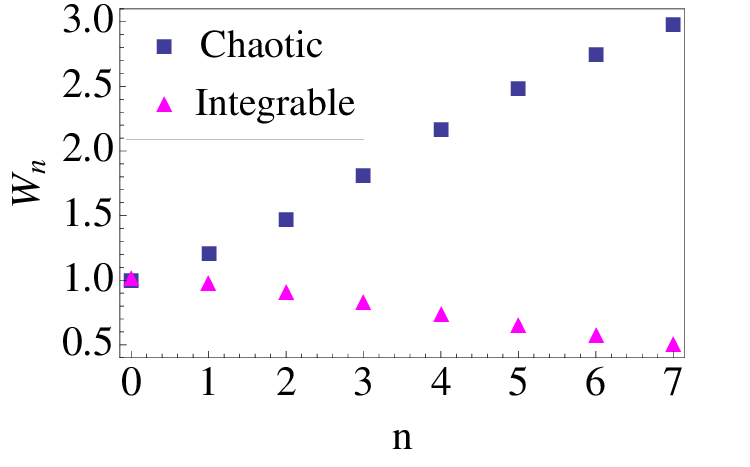,width=4.4cm  } }
\put(-1, 25){{ (a)}}
\put(42, 25){{ (b)}}

}
\end{picture}

\caption{ \label{stat} (Color online)   Statistical characteristics of the time series of $M_x^{(n)}(t)$ for the chaotic and the integrable systems: (a) ratios $R_n$ of the RMS value of $M_x^{(n)}(t)$  to the RMS value of $M_x^{(n-1)}(t)$; (b) relative width $W_n$ of the power spectrum $P^{(n)}(\omega)$ with respect to the width of $P(\omega)$.}
\end{figure}

We further propose two ways to quantify the above criterion. The first of them involves the root-mean-squared (RMS) values of $M_x^{(n)}(t)$, denoted as  $M_{rms}^{(n)}$. In Fig.~\ref{stat}(a), we plot the quantity
$R_n \equiv M_{rms}^{(n)}/M_{rms}^{(n-1)}$ as a function of $n$. For the nonchaotic time series, $R_n$ exhibits saturation, while for the chaotic time series it increases nearly linearly without apparent limit. The second way is to look at the evolution of $v_n$ defined as the square root of the variance for the positive-$\omega$ part of $P^{(n)}(\omega)$  [$v_0$ corresponds to $P(\omega)$]. In Fig.~\ref{stat}(a), we plot $W_n \equiv v_n/v_0$.
Here the difference is that, in the chaotic case, $W_n$ asymptotically increases with $n$, while in the nonchaotic case it decreases, asymptotically approaching zero.

\begin{figure} []

\ \ \ \ \ \ \ \ \ \ \ Chaotic \ \ \ \ \ \ \ \ \  \ \ \ \ \ \ \ \ \ \ \ \ \ \ \ \ \ \   Integrable

\includegraphics[width=4.2cm]{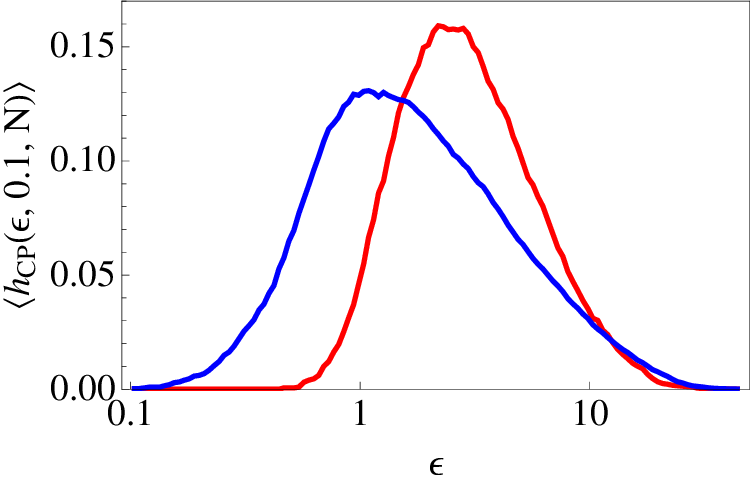} 
\includegraphics[width=4.2cm]{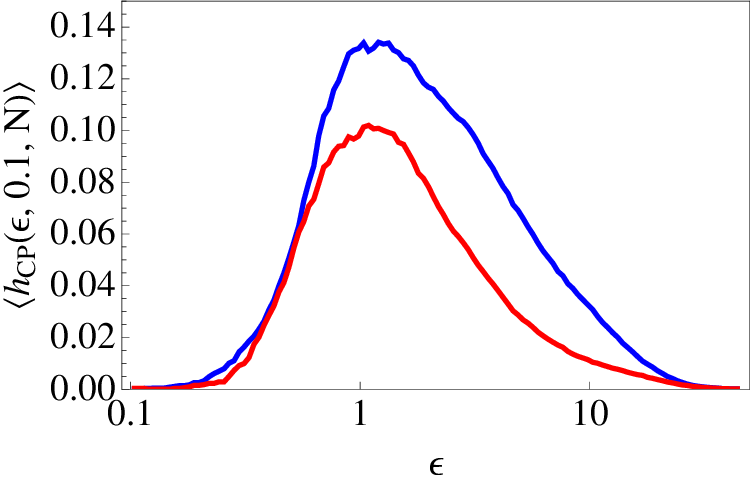}
\caption{(Color online) $\mathit{h}_{CP}(\epsilon, \tau,10)$ calculated for both $M_x(t)$ [blue] and $M_x^{(7)}(t)$ [red] as described in the text. The values of $\varepsilon$ for  $\mathit{h}_{CP}(\epsilon, \tau,10)$ associated with the 7th derivative were rescaled by dividing the true values by the following factors: $1.13 \cdot 10^{4}$ in the chaotic case and $1.31 \cdot 10^{3}$ in the integrable case. These factors were chosen such to make the spread between the maximum and the minimum values of $M_x^{(7)}(t)$ equal to that of the corresponding $M_x(t)$, thereby, in effect, compensating for the different dimensions of $M_x(t)$ and $M_x^{(7)}(t)$. }
\label{CP-fig}
\end{figure}

The fact that the higher-order derivatives look more random  for the chaotic time series also has quantifiable consequences in terms of the rate of the $(\varepsilon , \tau )$-entropy production. In particular, we notice some qualitative difference between the chaotic and the integrable cases, when we compare the CP $(\varepsilon, \tau)$-entropy rate for  the $7^\text{th}$  derivative with the CP $(\varepsilon ,\tau)$-entropy rate for the original time series. This comparison is presented in  Fig. \ref{CP-fig}. It indicates that  for the integrable system, the maximum  value of $\mathit{h}_{CP}(\epsilon, \tau,N)$ for fixed $\tau$ and $N$ is smaller for the higher-order time derivative than for the original time series, which implies that the derivative  produces less information. The situation is  opposite for the chaotic system.

\subsection{Noisy time series}
\label{Noisy}

In general, the effect of noise imposes a considerable limitation on the ability of the method introduced in the previous section to distinguish integrable from chaotic time series. In this section, we estimate the acceptable level of noise for the above method to work properly and present a practical example of how to deal with the white noise. 

Let us consider a chaotic system with an intrinsic power spectrum having an exponential tail of the form (\ref{p_of_omega}). Let us now add an additive white noise with power spectrum $P(\omega)=Q_0$ to the original time series. The cutoff frequency at which the noise intensity becomes comparable with the intrinsic exponential tail is given by $\omega_{c}=\frac{1}{\gamma} \log \frac{P_0}{Q_0}$. The frequency at which the power spectrum of the $\text{\it n}^\text{th}$ derivative exhibits a peak is given by $\omega_{max}=2n/\gamma$. The effect of the noise on our method will be tolerable as long as $\omega_{max}<\omega_c$, i.e., 
\begin{equation}
2n < \log(\frac{P_0}{Q_0}).
\label{ineq}
\end{equation}

In an experiment, $Q_0=\delta M^2 \delta t$, where $\delta M$ is the RMS amplitude of either an external physical noise or the noise due to finite accuracy of measurements, and $\delta t$ is the time resolution of measurements. Assuming also that $P_0 \sim P(0) \approx M_{rms}^2 \tau$, where $M_{rms}$ is the RMS of the time series  and $\tau$ is the characteristic time scale of the $M(t)$, we obtain from Eq. \ref{ineq} 
\begin{equation}
\delta M < M_{rms} ( \frac{\tau}{\delta t})^{0.5}e^{-n}.
\end{equation}

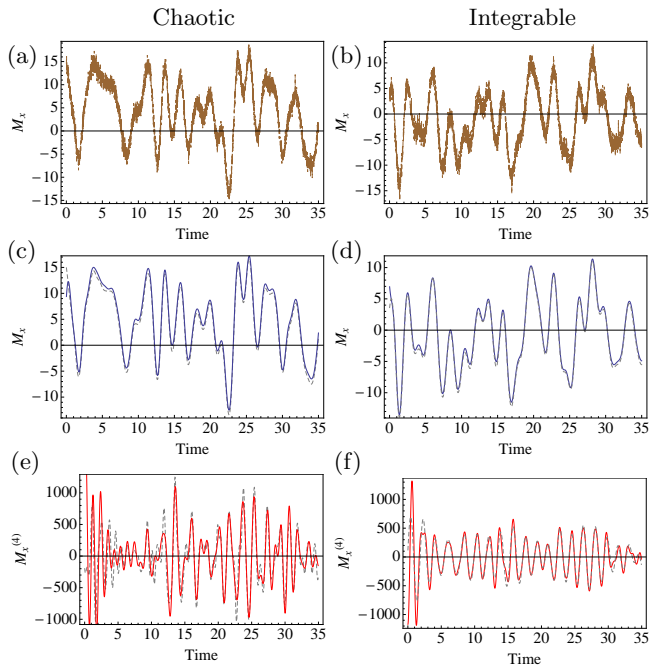
\begin{figure} \setlength{\unitlength}{0.1cm}

\begin{picture}(90 , 85 ) 
{

\put(0, 54){  \epsfig{file=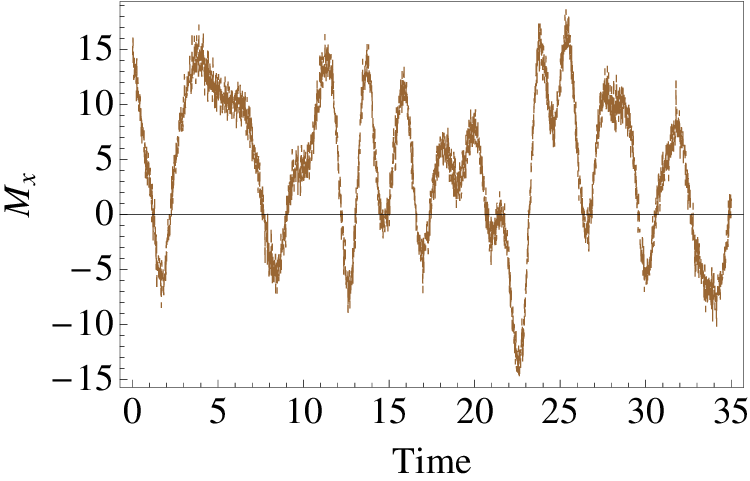,width=4.2cm } }
\put(43, 54){ \epsfig{file=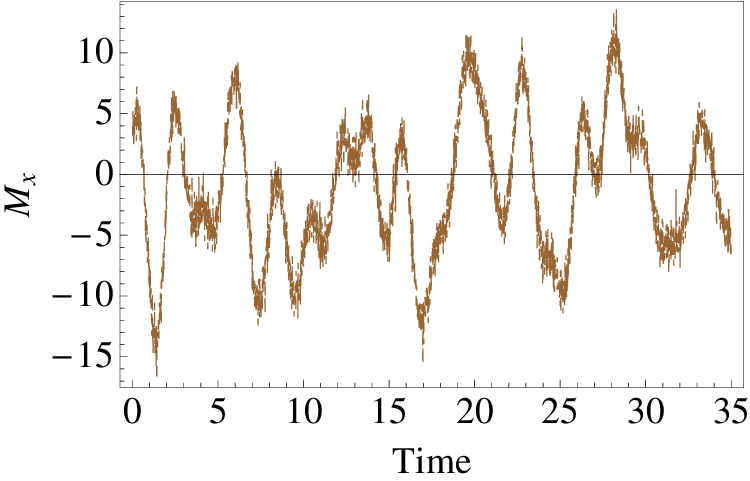,width=4.2cm  } }
\put(-0.5, 78){{ (a)}}
\put(42.5, 78){{ (b)}}

\put(0, 25.5){  \epsfig{file=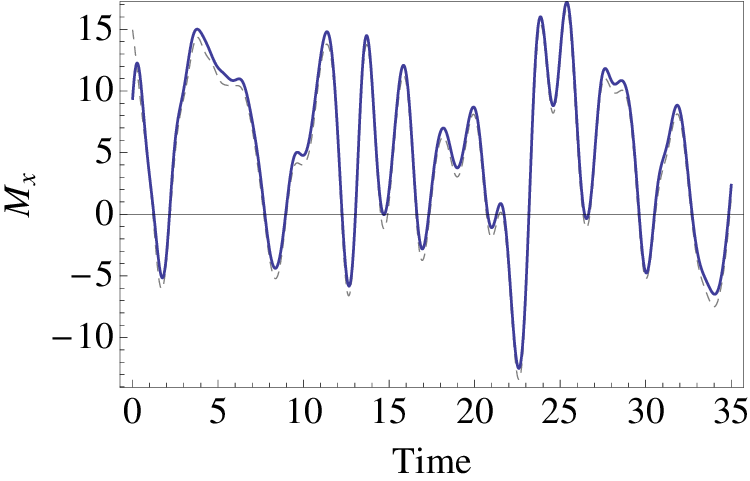,width=4.2cm } }
\put(43, 25.5){ \epsfig{file=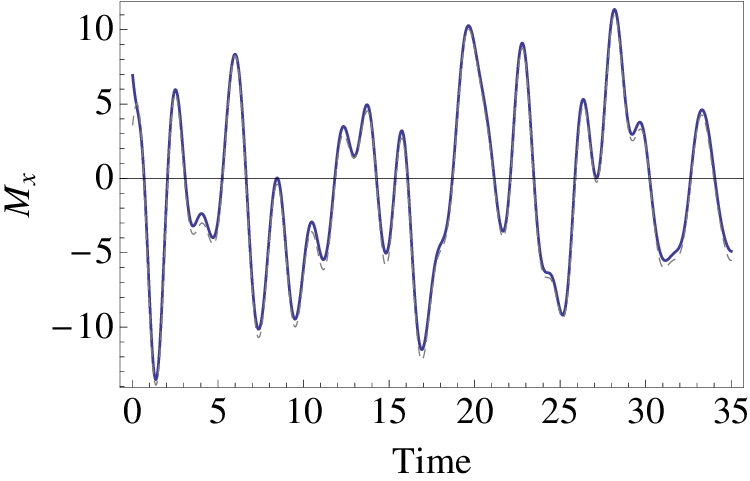,width=4.2cm  } }
\put(-0.5, 52){{ (c)}}
\put(42.5, 52){{ (d)}}
\put(0, -2){  \epsfig{file=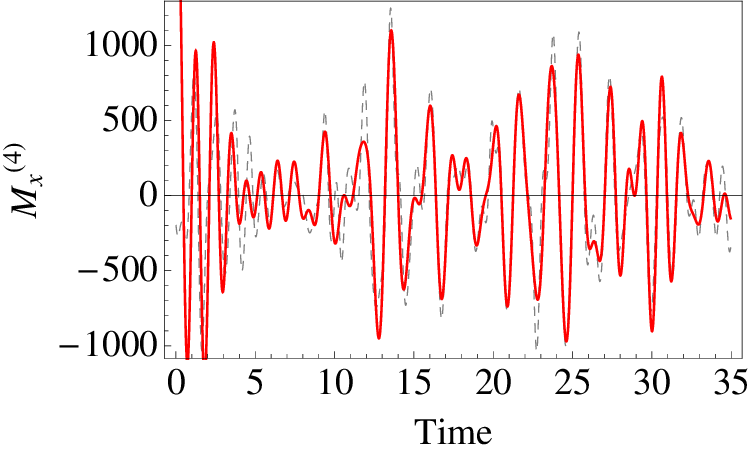,width=4.2cm } }
\put(43, -2.5){ \epsfig{file=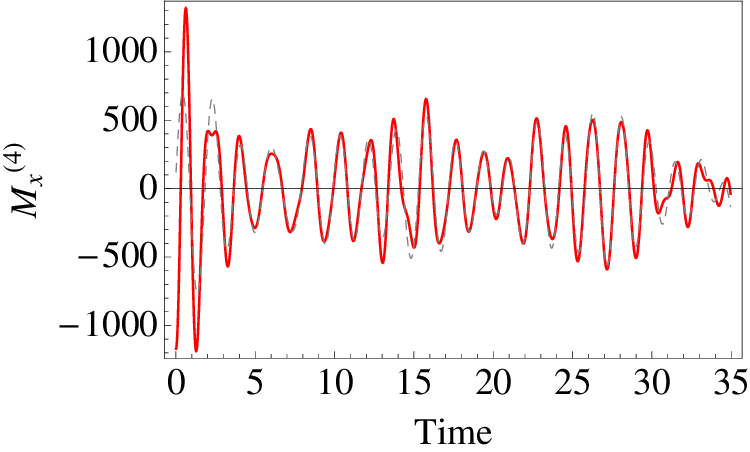,width=4.2cm  } }
\put(0, 24){{ (e)}}
\put(43, 24){{ (f)}}
\put(20,83){Chaotic}
\put(62,83){Integrable}
}
\end{picture}

\caption{ \label{noisy} (Color online) (a) and (b) represent noisy versions of the time series shown in Fig. \ref{main} (a) and (b) respectively. The notations are also the same as in Fig.~\ref{main}. (c) and (d) show the same time series after being filtered at the cutoff frequency 1 Hz (solid blue lines) and compare them to the original noise-free time series  (dashed gray lines). (e) and (f) show the fourth-order derivatives of the filtered time series (solid blue lines) and compare them to the  derivatives of the original noise-free time series ( dashed gray lines).  }
\end{figure}

When inequality (\ref{ineq}) is satisfied,  one possible way to deal with the noise is simply to filter it out by cutting the power spectrum at the critical frequency $\omega_c$.   In Fig. \ref{noisy}, we demonstrate that such a procedure indeed preserves  the  distinction between the higher-order time derivatives of chaotic and integrable time series.  Figures \ref{noisy} (a) and \ref{noisy} (b) represent noisy versions of the time series shown in Figs. \ref{main} (a) and (b) respectively. The noisy time series were obtained from the original ones by adding Gaussian white noise uncorrelated between adjacent time steps and characterized by  root-mean-squared amplitude equal  approximately to $1/8 $ of the root-mean-squared amplitude of the original signal.   The resulting value for $Q_0$ was such that  $P_0/Q_0\approx10^{-6}$, so that $\omega_c=1 \text{Hz}$. Figs. \ref{noisy} (c-f) illustrate how the  filtering of the noisy time series before taking the time derivatives remedies the effect of the noise. In Figs. \ref{noisy} (c-d), we compare the original time series without noise and the filtered noisy time series. In Figs. \ref{noisy} (e-f), we do the same for the fourth-order time derivatives.


\subsection{Short time series}
\label{Short}

\begin{figure}[h] \setlength{\unitlength}{0.1cm}

\begin{picture}(90 , 85 ) 
{

\put(0, 54){  \epsfig{file=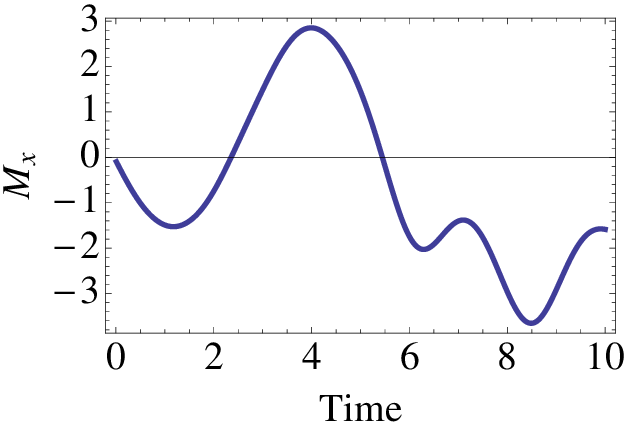,width=4.2cm } }
\put(43, 54){ \epsfig{file=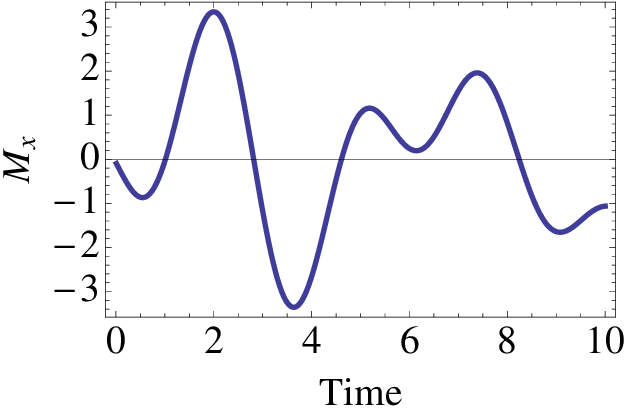,width=4.2cm  } }
\put(-0.5, 78){{ (a)}}
\put(42.5, 78){{ (b)}}

\put(0, 27){  \epsfig{file=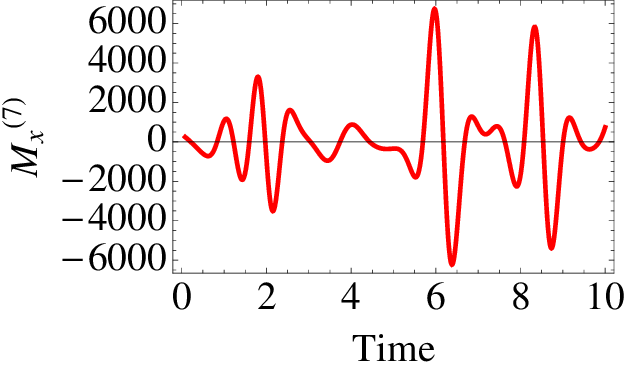,width=4.2cm } }
\put(43, 27){ \epsfig{file=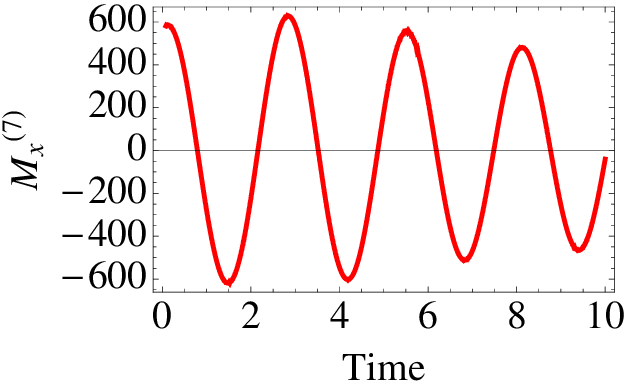,width=4.2cm  } }
\put(-0.5, 52){{ (c)}}
\put(42.5, 52){{ (d)}}
\put(0, -2){  \epsfig{file=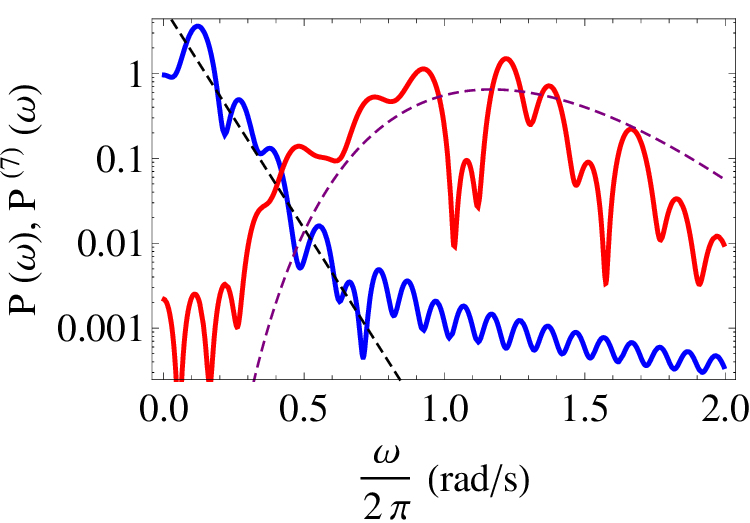,width=4.2cm } }
\put(43, -2.5){ \epsfig{file=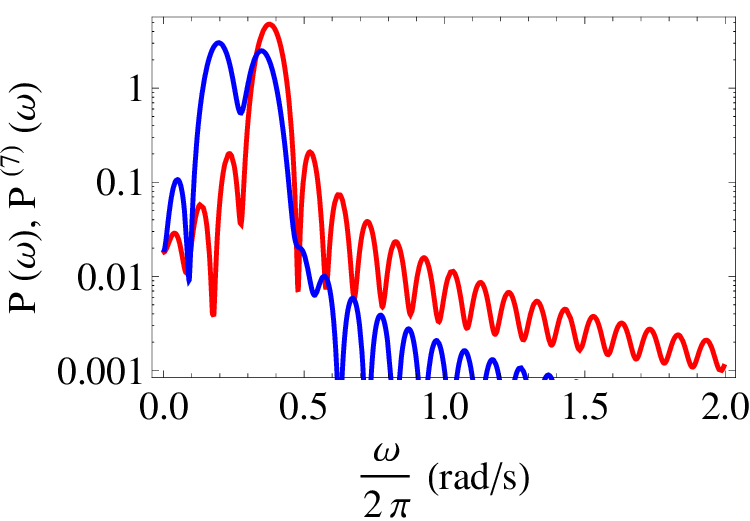,width=4.2cm  } }
\put(0, 24){{ (e)}}
\put(43, 24){{ (f)}}
\put(20,83){Chaotic}
\put(62,83){Integrable}
}
\end{picture}

\caption{ \label{short} (Color online) Time series of $M_x(t)$ for  $4 \times 4$ spin lattices. The notations and the interactions constants are the same as in Fig.~\ref{main}. Frames (a,b,c,d) now represent the entire time series used to obtain frames (e,f). }
\end{figure}
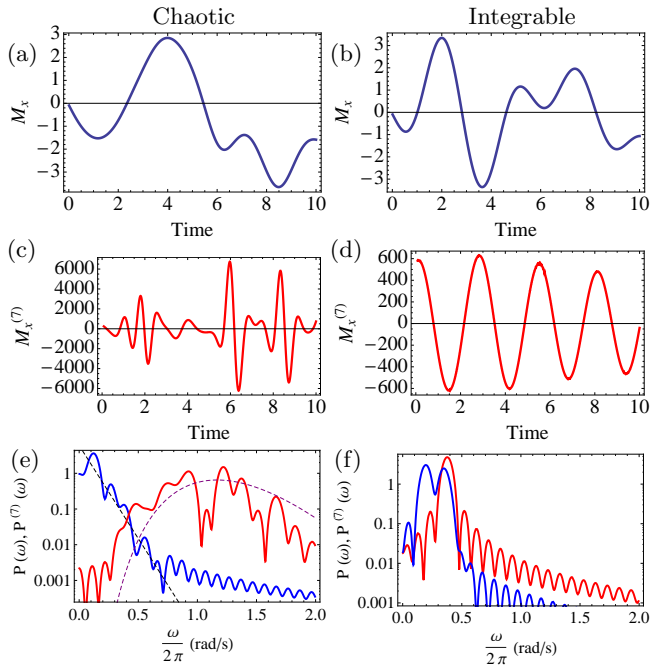

In the power spectrum of the nonchaotic time series shown in Fig.~\ref{main}(f), one can distinguish some discrete frequencies, which is already an indication of integrability. 
The chaos indicators extracted from the derivatives of the power spectrum are mainly intended for the situations when this discreteness is not discernible due to either too large number of discrete frequencies or too short length of the time series. 
 In Fig. \ref{short}, we illustrate the effectiveness of our method by applying it to a very short time series of length $T=10$ produced by two $4 \times 4$ square lattice spin clusters with the same nearest neighbor coupling coefficients as the previously considered two clusters. Detecting chaos from such a short time series using just the power spectrum of $M_x(t)$ is difficult, because it is contaminated by the oscillations and the power-law decays associated with the short length of the time series.  However, taking the 7th derivative reveals the expected qualitative difference between the chaotic and integrable systems and the good quality  fit of form (\ref{criterion}). 


\section{Transition to integrability in classical spin systems}
\label{transition}

In this section, we illustrate how the exponential tail of the power spectrum behaves during the transition from chaotic dynamics associated with generic nonintegrable Hamiltonian to the nonchaotic dynamics associated with the Ising Hamiltonian.

 It was shown recently \cite{fine-12} that the maximum Lyapunov exponents of classical spin lattices exhibit a power-law decrease to zero as the system's Hamiltonian approaches the Ising limit. One could have expected a similar behavior for $1/\gamma $ in Eq.(\ref{p_of_omega}). However, we  found out that it is not $1/\gamma $ that decreases, but rather the prefactor $P_0$. 
 
\begin{figure} \setlength{\unitlength}{0.1cm}

\begin{picture}(88 , 46 ) 
{
\put(0, 0){  \epsfig{file=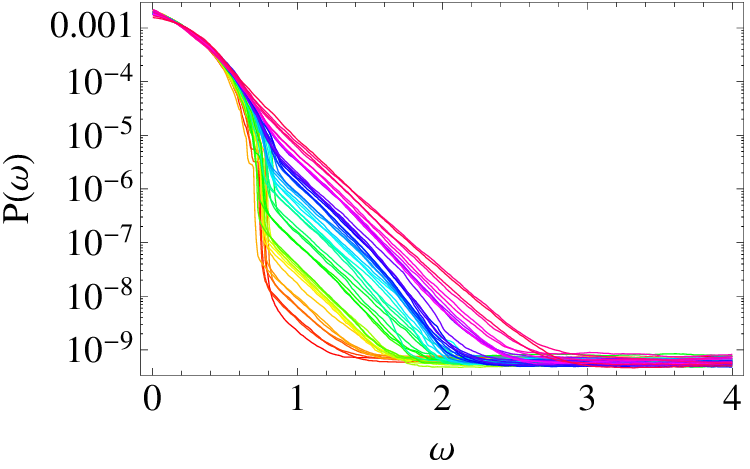,width=7.5cm } }
\put(37, 23){  \epsfig{file=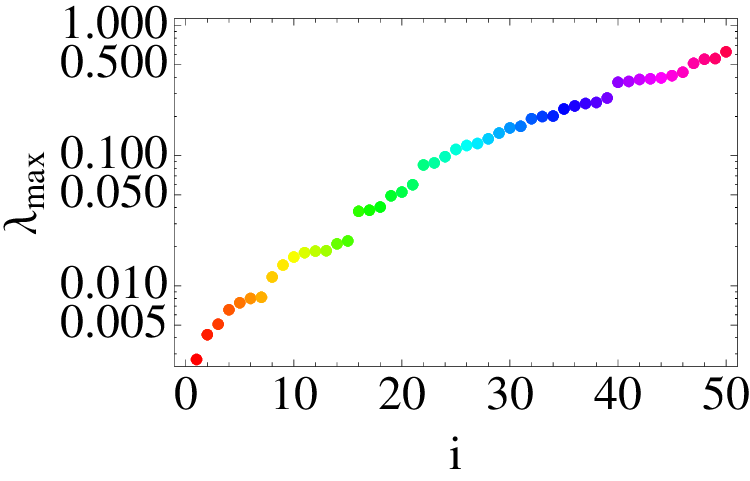,width=3.5cm } }
}
\end{picture} 

\caption{ \label{spectra}(Color online) Power spectra for 50  realizations of $J_x,\ J_y,\ J_z$   in the vicinity of the Ising limit subject to the constraint $J_x^2+J_y^2+J_z^2=1$ for a cubic spin lattice consisting of $16\times16\times16$ spins.  The specific realizations of  $J_x,\ J_y,\ J_z$ were selected from the data set used in Ref. \cite{fine-12}. The corresponding values of $\lambda_{max}$ decrease monotonically by more than two orders of magnitude from the upper power spectra to the lower power spectra. The values of $\lambda_\text{max}$ themselves are shown in the inset.  They were computed in Ref. \cite{fine-12}.}
\end{figure}

 The above result is presented in Fig. \ref{spectra}, where we plot power spectra for a family of Hamiltonians gradually approaching the Ising limit. 
These Hamiltonians were selected from a larger set of Hamiltonians used in Ref. \cite{fine-12} for the survey of the largest Lyapunov exponents. The values of $\lambda_{max}$ corresponding to the power spectra in Fig. \ref{spectra} are given in the inset of that figure. These values  cover more than two orders of magnitude. The smaller $\lambda_\text{max}$, the closer the Hamiltonian to the Ising limit.  It can be seen in Fig. \ref{spectra} that the values of $\gamma$ remain nearly the same during the approach to the Ising limit, while $P_0$ decreases by approximately four  orders of magnitude.

\section{The Toda lattice}
\label{Toda}

We  found a nongeneric completely integrable many particle classical system that exhibits an exponential tail in its power spectrum , namely, the Toda lattice \cite{Toda-89}. The Toda lattice consists of $N$ particles in one dimension characterized by coordinates $q_i$ and interacting according to the potential $V(q_i,q_{i+1})=c\ e^{-\alpha(q_i-q_{i+1})}$, where $c$ and $\alpha$ are constants.

 We numerically investigated a Toda lattice  consisting of 32 particles with periodic boundary conditions and  $c=1$ and $\alpha=1$. In this case, the equations of motion are
  \begin{eqnarray}
&  \dot{q_i}=p_i,\\
& \dot{p_i}=e^{q_i-q_{i-1}}-e^{q_{i+1}-q_i}.
  \end{eqnarray}

\begin{figure} \setlength{\unitlength}{0.1cm}

\begin{picture}(90 , 55 ) 
{

\put(-0.5, 50){{\small (a)}}
\put(43., 50){{\small (b)}}
\put(0, 25.5){  \epsfig{file=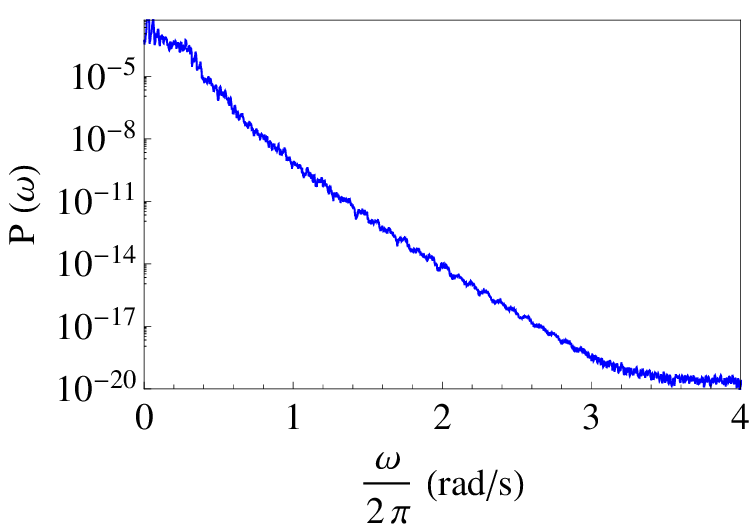,width=4.2cm } }
\put(43, 25.5){ \epsfig{file=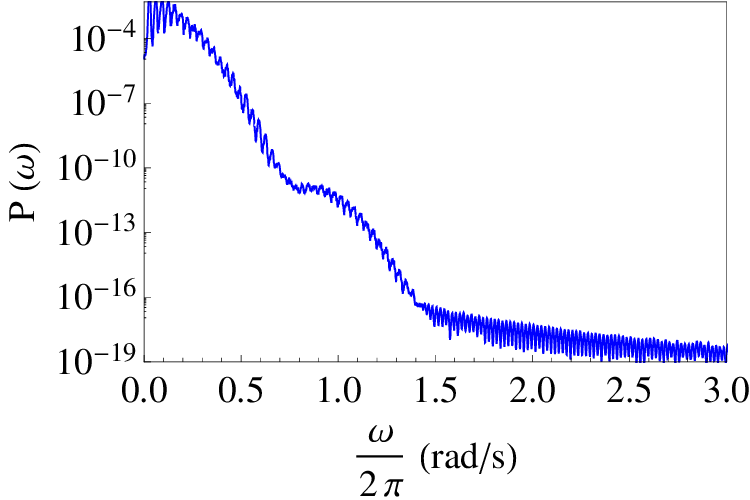,width=4.2cm  } }
\put(-0.5, 22){{\small (c)}}
\put(42.5, 22){{\small (d)}}
\put(0, -2){  \epsfig{file=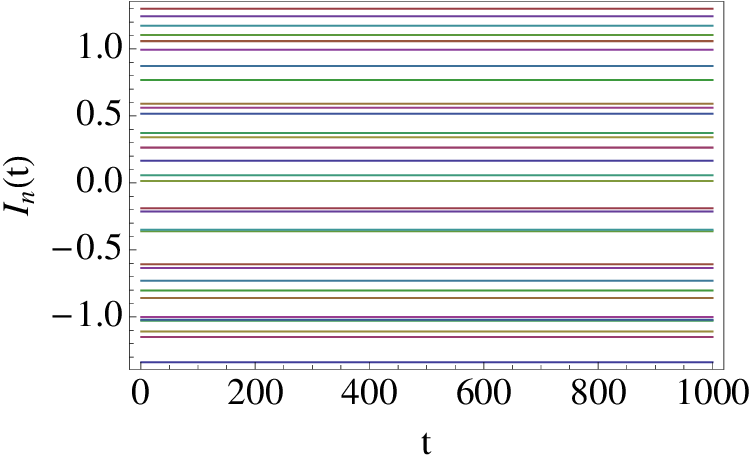,width=4.2cm } }
\put(42, -2.5){ \epsfig{file=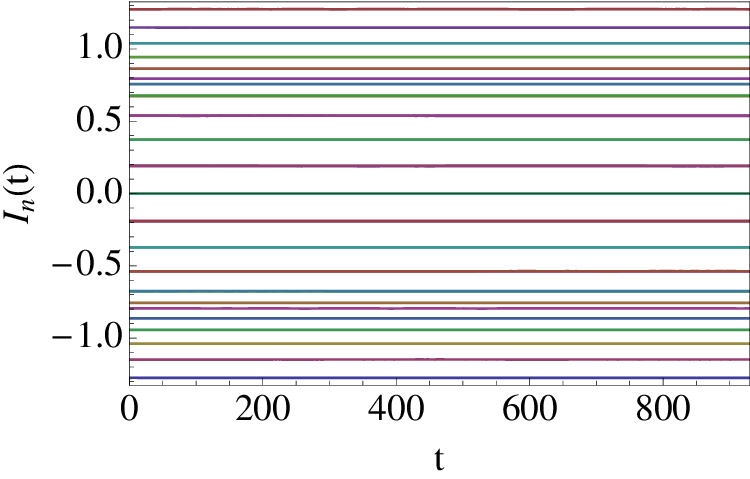,width=4.2cm  } }
}
\end{picture}

\caption{ \label{toda} (a)  The power spectra of a Toda lattice consisting of 32 particles with random initial conditions. (b) Same for the initial conditions chosen according to Eq. \ref{toda-sin}. (c) and (d) represent the integrals of motion $I_n$ as a function of time for the initial conditions of (a) and (b) respectively.   }
\end{figure}
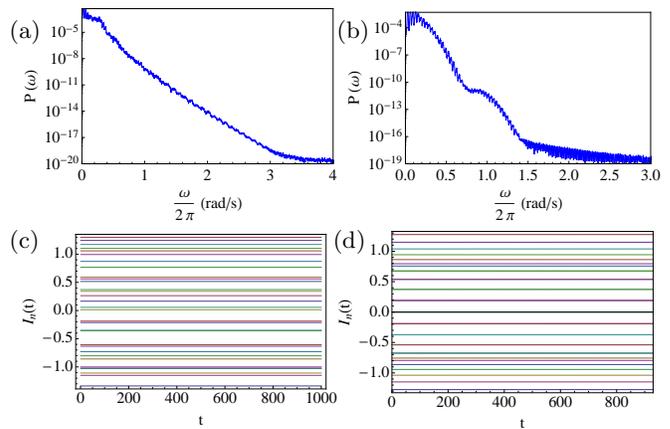

We computed the power spectrum for the time series of the coordinate of a single particle for two sets of initial conditions. The first set corresponds to the random choice of initial coordinates and momenta, subject to the constraint that the total momentum equals zero. The second set of  initial conditions is given by
\begin{equation}
q_i(0)=A \sin \left( \frac{2\pi  i}{N} \right), \ \ \ i=1,.\ .\ .\ N
\label{toda-sin}
\end{equation}
and $p_i(0)=0$.
The power spectra for the two time series are depicted in Figs. \ref{toda} (a) and \ref{toda} (b)  respectively. In the first case, the power spectrum has a nearly exponential tail at high frequencies, while in the second case it does not exhibit a pure exponential decay. 

 We analyzed the integrals  of motion $I_n$ in both cases. These integrals can be defined as the eigenvalues of the matrix $L$ given by
\begin{equation}
 L=\begin{bmatrix}
b_1 &a_1  &0 &\ldots  &a_n \\ 
a_1 &b_2  &a_2 &  & \\ 
0&a_2&b_2&\ddots&\\
\vdots &  &\ddots  &\ddots&a_{n-2} \\ 
 a_n& &  &a_{n-1}  &b_n 
\end{bmatrix}
\label{L-matrix}
\end{equation}
where   $a_i=e^{\frac{q_{i+1}-q_i}{2}}$ and  $b_i=p_i,$ \cite{flaschka}.  We notice in Figs. \ref{toda} (c) and \ref{toda} (d) that $I_n$ do not depend on time as expected which verifies the accuracy of our numerical simulation. Secondly, we notice that $I_n$ are randomly distributed in the first case, while they are more uniformly distributed in the second case. The dependence of $I_n$ on the initial conditions and the possible similarity to the repulsion of energy levels in quantum systems is an interesting issue which extends beyond the scope of the present article. 
In Appendix C, we  also report our brief investigations of a Toda lattice with a truncated exponential potential.

Even though the power spectrum in Fig. \ref{toda} (a)  exhibits a nearly exponential tail, we suspect that such a behavior is not typical for integrable systems. The Toda lattice is an exceptional integrable model, because, despite being integrable, it is not separable \cite{gutzwiller}. In contrast, the Ising model for classical spins is separable and thus, in our opinion, represents the typical behavior of an integrable many-particle system.
 We  have not tried to formulate a constraint on the integrable Hamiltonians that would exclude the exponential tails of the power spectra. 

\section{Quantum spin systems}
\label{quantum}

It is natural to ask whether exponential tails of power spectra can discriminate  integrable from nonintegrable quantum systems.  If yes, can this be done on the basis of  the time series analysis of a single macroscopic observable? In this section, we address the above issue by investigating the power spectra of integrable and nonintegrable quantum spin systems. 

 One difference between quantum and classical time series is that the spectrum of the former consists of discrete frequencies for both integrable and nonintegrable systems. Nevertheless, for large quantum systems, the power spectra become effectively continuous because the spacings between the energy levels become exponentially small, and the timescale on which the discretness of the spectrum is observable becomes exponentially long. 

Experimentally, the power spectra for quantum spin systems can be obtained with the help of nuclear magnetic resonance (NMR) by performing  the Fourier transform of the free induction decay signal \cite{abragam}. The tails of thus obtained NMR power spectra of  CaF$_2$ were found in Ref. \cite{lundin}  to have exponential shape. This suggest that the signatures of chaos in terms of higher-order time derivatives of the magnetization time series may be obtainable from the direct monitoring of the equilibrium fluctuations of nuclear spin polarization in solids. Such fluctuations were measured by several NMR groups \cite{sleator,schlagnitweit} although not yet with high enough accuracy.  From the theoretical perspective,  our recent work \cite{Elsayed-12} indicates that the time series for the expectation values of the quantum mechanical operator of the total nuclear magnetization in a pure quantum state also has the power spectrum given by the Fourier transform of the NMR free induction decay.  

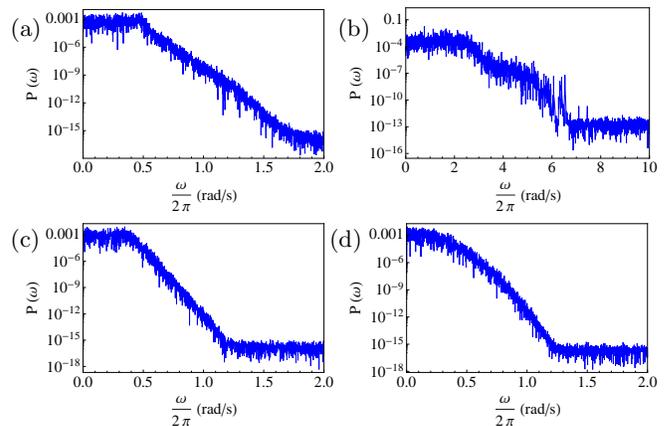
\begin{figure} \setlength{\unitlength}{0.1cm}

\begin{picture}(90 , 58) 
{

\put(0, 28.5){  \epsfig{file=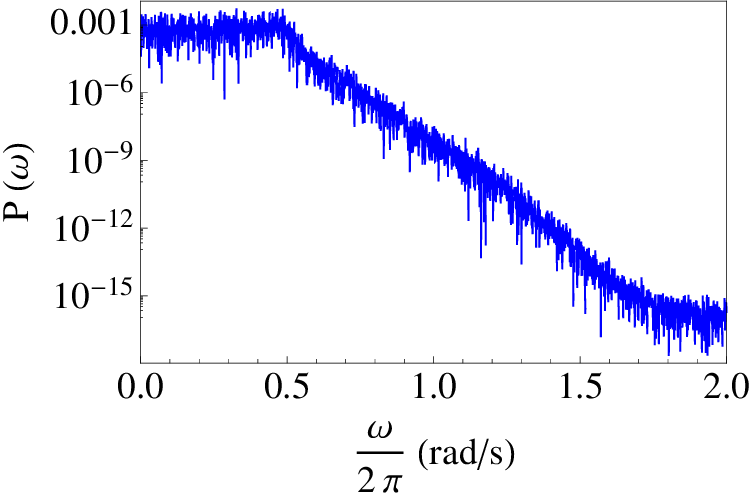,width=4.1cm} }
\put(43, 28.5){ \epsfig{file=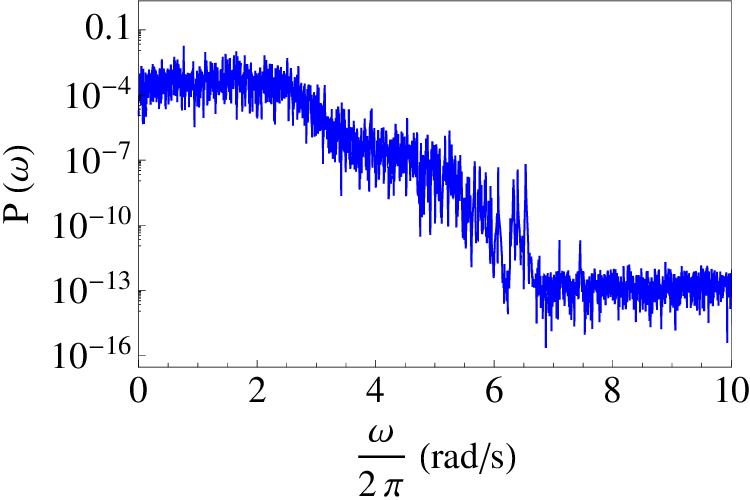,width=4.1cm } }

\put(43, 0){  \epsfig{file=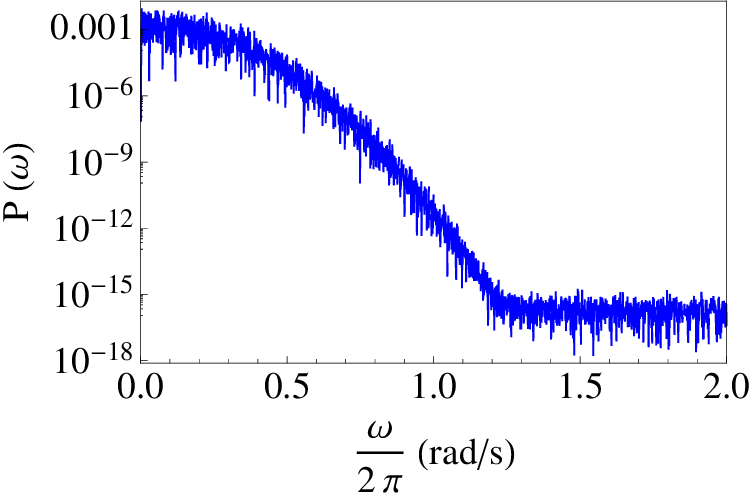,width=4.1cm} }
\put(0, 0){ \epsfig{file=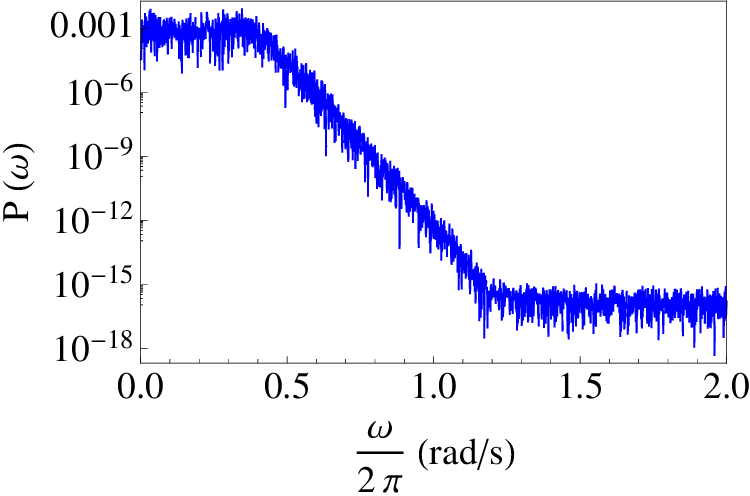,width=4.1cm  } }
\put(-2, 24){{ \small(c)}}
\put(41.5, 24){{\small (d)}}

\put(-2, 52){{ \small(a)}}
\put(41.5, 52){{ \small(b)}}

}
\end{picture} 

\caption{ \label{Qspectra}  Power spectra for different quantum spin 1/2 lattices computed from the time series representing the expectation values for the operator $ \hat{M_x} $.  The lengths of the time series are greater than 2000. The initial states are selected randomly from the respective infinite-temperature ensembles.  The systems are: (a) Nonintegrable 16-spin chain with nearest neighbor coupling coefficients $J_x=0.267, \ J_y=0.535$ and $J_z=-0.802$ and next-nearest-neighbor coupling of strength 30\% of the nearest-neighbor coupling. (b) Square $5\times 4$ lattice  with XXZ coupling coefficients $J_x=J_y=0.433$ and $J_z=-0.356$  (c) Integrable  XXZ chain consisting of 16 spins with coupling coefficients $J_x=J_y=0.612$ and $J_z=-0.5$. (d) Integrable XX chain consisting of 16 spins with coupling coefficients $J_x=J_y=0.707$ and $J_z=0$.} 

\end{figure}

In Fig. \ref{Qspectra}, we present the power spectra of four quantum spin 1/2 lattices, integrable and nonintegrable, computed from the time series of the expectation value of the total magnetization operator $M_x$ at the infinite-temperature equilibrium. The Hamiltonians of these lattices have the form given by Eq. \ref{hamiltonian} with $S_x$, $S_y$, and $S_z$ denoting the quantum spin 1/2 operators.

 

   
  Figs. \ref{Qspectra} (a) and \ref{Qspectra} (b) correspond to two nonintegrable lattices.   The first case is an anisotropic Heisenberg spin chain consisting of 16 spins with  $J_x=0.267, \ J_y=0.535$ and $J_z=-0.802$ and next nearest neighbor coupling of strength 30\% of the nearest neighbor coupling. We see clearly an exponential tail in its spectrum. The second case is a two-dimensional  $5\times 4$ quantum spin lattice  with XXZ coupling coefficients ($J_x=J_y=0.433$ and $J_z=-0.356$). Here the power spectrum is not purely exponential. This case may require further investigation to exclude finite-size effects.
  
 In Fig. \ref{Qspectra} (c) and Fig. \ref{Qspectra} (d), we show the power spectra  for two integrable models. The first model is a Bethe-ansatz  XXZ  integrable spin chain consisting of 16 spins with coupling coefficients $J_x=J_y=0.612$ and $J_z=-0.5$.  We notice that its power spectrum has an exponential tail. Such a behavior representative of classical chaotic systems is not entirely surprising in this case given that in Ref. \cite{fine-04} it was found that this chain exhibit another generic signature of chaos, namely, the exponential tails of time correlation functions.  The second case is an XX spin chain consisting of 16 spins with  coupling coefficients $J_x=J_y=0.707$ and $J_z=0$. We observe that the power spectrum of this model is a Gaussian function of frequency. This result is expected since the correlation function $\langle M_x(t)M_x \rangle$ is known to be Gaussian \cite{stolze95}. 

The above numerical results imply that making a  clear distinction between the power spectra of integrable and nonintegrable quantum systems may be a challenging task, even though, in view of NMR spectra of CaF$_2$, an exponential tail of the spectrum may still be a generic feature of nonintegrable quantum systems. This is consistent with the broader subtlety of the interplay between the onset of chaos and the classical-to-quantum transition (see, e.g., Refs.\cite{Kapulkin-08,Finn-09,Kingsbury-09}). We further remark in this regard that our very recent investigations indicate that  nonintegrable lattices of spins 1/2 do not exhibit exponential sensitivity to small perturbations unlike their classical counterparts \cite{Fine-14}. 


\section{Concluding remarks}
\label{discussion}

 In this article, we have demonstrated that examining the behavior of higher-order time derivatives of the time series of total magnetization is an effective way to distinguish between  chaotic and integrable classical spin systems. The utility of the above method  is likely to be limited by the experimental noise \cite{sigeti-pre-95}, but this  limitation can be partially overcome with the help of an appropriate high-frequency filtering described in Section~\ref{Noisy}. The influence of the noise can be further decreased by measuring fluctuations of total magnetization for larger samples.

The above criterion of chaos is related to the presence or the absence of exponential high-frequency tails in the power spectra corresponding to the time series considered.
Our findings indicate that microscopic chaos in systems of classical spins at infinite temperature generically implies  exponential tails of power spectra. The reverse is also very likely to be true given the results of Ref. \cite{fine-12}, which indicate that the Ising Hamiltonian is the only integrable nearest-neighbor Hamiltonian for classical spin lattices, while our findings clearly show that the power spectra corresponding to the Ising Hamiltonian do not have exponential tails. We further expect that our chaos-detection method remains valid for classical spin lattices with longer-range interactions. In this case, no survey of the largest Lyapunov exponents analogous to that of Ref. \cite{fine-12} has been performed yet, but it can be reasonably expected that only Ising Hamiltonians with various long-range interactions remain integrable. In principle, however, one should be able to artificially adjust Ising coupling constants to make the resulting power spectrum acquire apparent exponential tail.   

We now discuss the extensions of the above criterion beyond infinite temperature, beyond the magnetization time series, to quantum spin systems and to classical non-spin systems.

We expect that the above criterion remains applicable to spin lattices at high enough finite temperatures. However, as the temperatures become lower the criterion needs further verification, in particular, in the proximity of magnetic phase transitions and below. The concern here is that, if large parts of the system exhibit slow non-ergodic behavior (for example, fluctuations of different magnetic domains), then such a behavior may lead to the resulting power spectrum becoming decomposable into contributions from different parts of the system with different exponential tails, which, in turn, may lead to the overal nonexponential shape of the tails. 

In this article, we mostly concentrated on the time series of an extensive macroscopic observable --- the total magnetization of a spin lattice. One may wonder whether our results extend to (i) local observables, such as  individual spin coordinates, and (ii) higher powers of the total magnetization such as $ M_x^2(t)$. We expect that,  case (i), the chaotic character of spin dynamics still leads to the exponential tails of the power spectra, but then there exist much simpler ways to discriminate generic Hamiltonians from the Ising Hamiltonian, because, in the Ising case, the one-spin motion is manifestly periodic. In case (ii), the detection of the difference between chaotic and nonchaotic systems in the time series of the higher powers of the total magnetization would encounter serious practical complication, due to the fact that the corresponding power spectra represent multiple convolutions of the original magnetization power spectrum. As a result, both the cutoff of the power spectra in an integrable system and the onset of the exponential tail in a chaotic system become shifted to higher frequencies and lower intensities, and hence become more difficult to observe.  The above considerations imply that the method of identifying chaos on the basis of the analysis of higher-order time derivatives has most value for analyzing the time series of extensive (i.e. additive) quantities such as the total magnetization.

The extension of the above chaos criterion to quantum spins 1/2 is less clear at the moment. As mentioned in Section~\ref{quantum}, the power spectra extracted from NMR experiments do exhibit exponential tails. However, the results of our own numerical investigations do not convey a consistent picture as far as the exponential tails, possibly because of the finite-size effects.

Regarding classical non-spin systems with many degrees of freedom, we expect  our criterion of chaos to be applicable to time series of fast fluctuating variables (as opposed to slow hydrodynamic variables) in classical models of simple liquids  at high enough temperatures (i.e. far enough from the crossover to the glassy behavior). 

We should further mention that our findings are consistent with those of Refs. \cite{frisch-81,farmer-82,sigeti-pre-95,sigeti1995} and Ref. \cite{tarek-thesis} for chaotic classical systems with a few degrees of freedom. 
In Ref. \cite{frisch-81}, the existence of an exponential tail in the spectra of chaotic time series was attributed  to the complex time singularities which exist when the equations of motion are analytically continued to the complex time plane. The positions of these singularities in the real time axis correspond to the positions of intermittent bursts of deep randomness in the time series whose amplitudes depend exponentially on the imaginary part of the complex time coordinate of the singularities \cite{frisch-81}. The decay constant of the exponential tail of the power spectrum is then determined by the singularities closest to the real time axis. 

Overall, as mentioned in the introduction, it is probably unavoidable, that, for any generic criterion of chaos not based of the direct observation of Lyapunov instabilities, one can propose artificially constructed integrable counterexamples. We have found two such counterexamples: the completely integrable classical Toda lattice and the Bethe-ansatz-integrable spin 1/2 chains. Both show signatures of an exponential tail in their power spectra. However, both of these systems do not allow separation of variables and thus might not be representative of the generic behavior of integrable systems.

Our investigation of the transition to the integrabily in classical spin systems presented in Fig.~\ref{spectra} indicates that the log-scale slopes of the high-frequency tails of the power spectra do not correlate with the values of Lyapunov exponents. This, in turn, possibly suggests that it is effective ergodicity of the single particle motion rather than Lyapunov instability as such that is important for the onset of the exponential high-frequency tails of the power spectra. Here  ``effective ergodicity'' means that the trajectories of different particles in the system observed over sufficiently long time lead to the same probability distributions in one-particle phase spaces. We further speculate that the systems not exibiting exponential high-frequency tails of the power spectra should be suspect of lacking ergodicity, in the sense that different parts of these systems produce different kinds of chaotic or nonchaotic motions, which make independent contributions to the tails of the power spectra, which, in turn may distort the overall exponential shape of the tail even when the individual contributions are exponential.

\acknowledgements

We are grateful to H. Kantz, P.Gaspard and D. Shepelyansky for discussions. We also thank A. S. de Wijn for providing the values of $\lambda_{max}$ presented in Fig. \ref{spectra}.
\\

\appendix
\section{Additional time derivatives for the example presented in Fig.~\ref{main}}
\label{additonal}

In Fig. \ref{main}, we presented the fragments of two time series $M_x(t)$ of the total spin polarization for the chaotic and the integrable $6 \times 6 \times 6$  spin clusters together with their respective 7th derivatives, $M_x^{(7)}(t)$. In Fig. \ref{derivatives} below, we show the plots for all the derivatives $M_x^{(n)}(t)$ of the same two time series up to the 9th order. There are several methods to efficiently compute the  derivatives of measured time series \cite{abel07}. We used the standard finite-difference numerical procedure for computing these derivatives. Namely, if the discretized time series in a given order was $ \left\{ ...,  \left( t_i, M_{x, i}^{(n)} \right), \left( t_{i+1}, M_{x, i+1}^{(n)} \right), \left( t_{i+2}, M_{x, i+2}^{(n)} \right), ... \right\}$, then the next-order derivative was obtained as 
$ \left\{ ...,  
\left( t_i, { M_{x, i+1}^{(n)}- M_{x, i}^{(n)} \over t_{i+1} - t_{i} } \right), 
\left( t_{i+1}, { M_{x, i+2}^{(n)}- M_{x, i+1}^{(n)} \over t_{i+2} - t_{i+1}}  \right) ,
 ... \right\}$.
 
\begin{widetext}
\begin{figure*} \setlength{\unitlength}{0.1cm}

\begin{picture}(190 , 140 )   
{

\put(0, 109){  \epsfig{file=plnon0.eps,width=4.2cm } }
\put(43, 109){  \epsfig{file=plising0.eps,width=4.2cm } }

\put(0, 82){  \epsfig{file=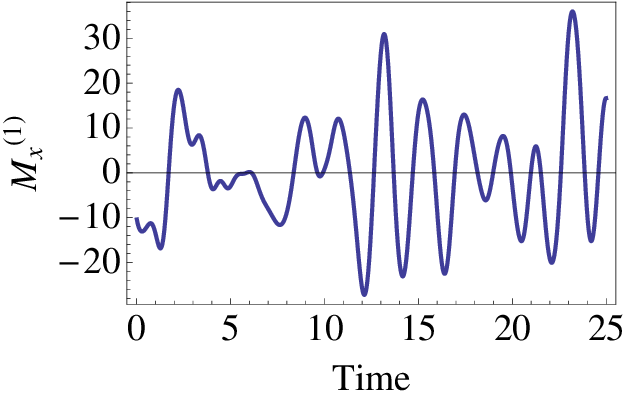,width=4.2cm } }
\put(43, 82){  \epsfig{file=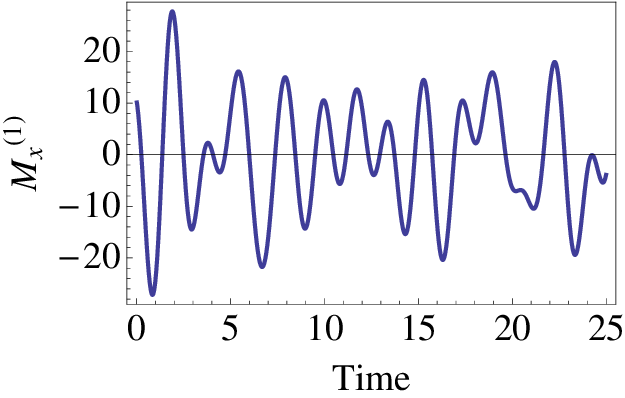,width=4.2cm } }

\put(0, 55){  \epsfig{file=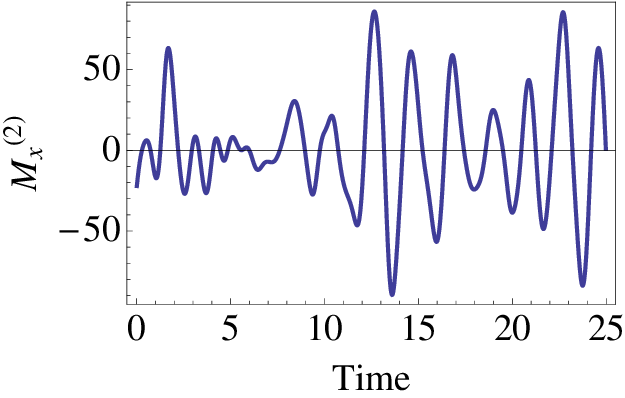,width=4.2cm } }
\put(43, 55){  \epsfig{file=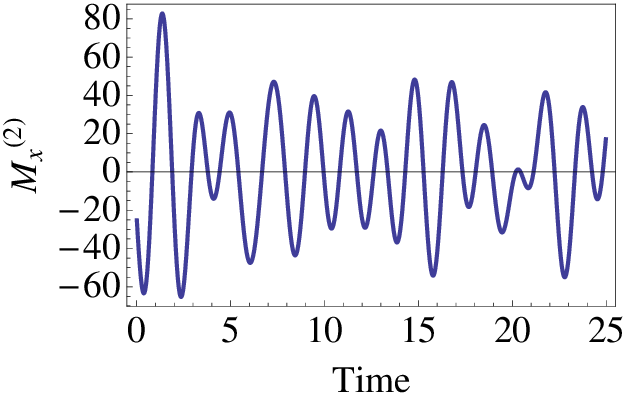,width=4.2cm } }

\put(0, 28){  \epsfig{file=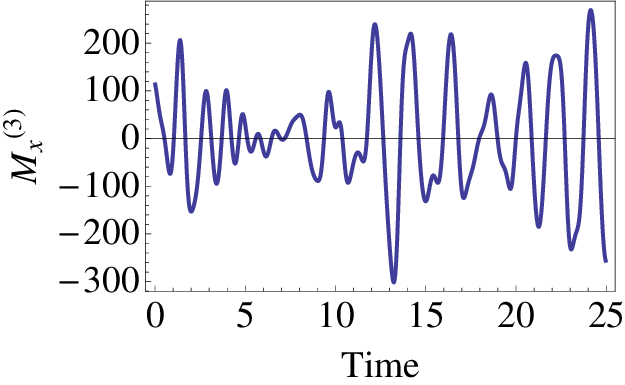,width=4.2cm } }
\put(43, 28){  \epsfig{file=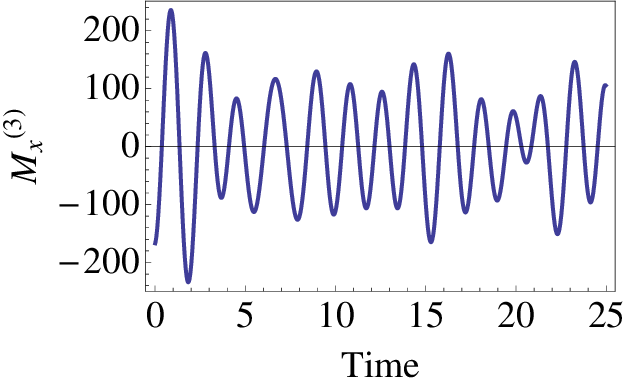,width=4.2cm } }

\put(0, 1){  \epsfig{file=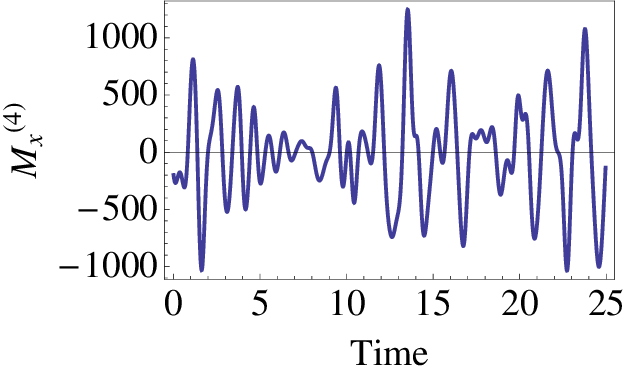,width=4.2cm } }
\put(43, 1){  \epsfig{file=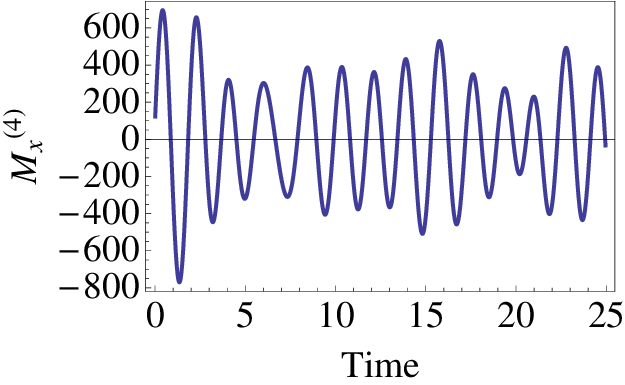,width=4.2cm } }

\put(19,138){Chaotic}
\put(62,138){Integrable}

\put(100, 109.5){  \epsfig{file=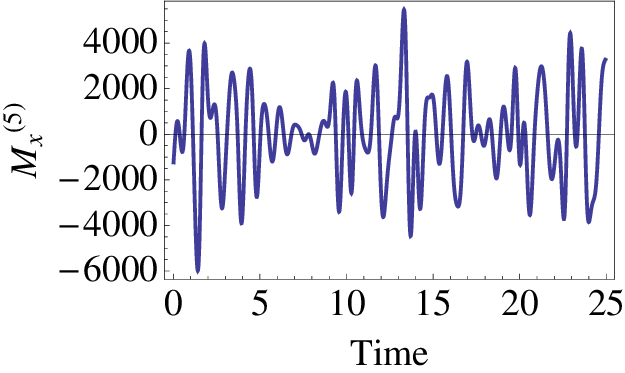,width=4.2cm } }
\put(143, 109.5){  \epsfig{file=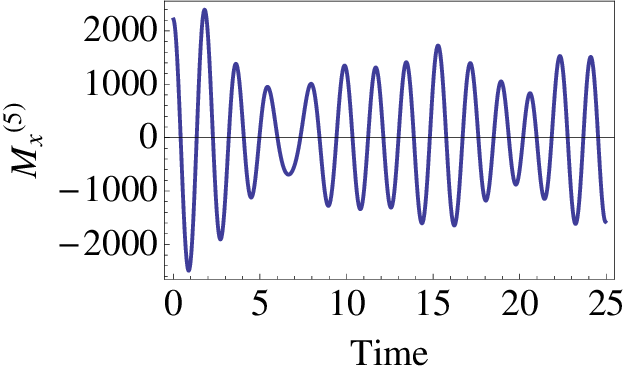,width=4.2cm } }

\put(100, 82.5){  \epsfig{file=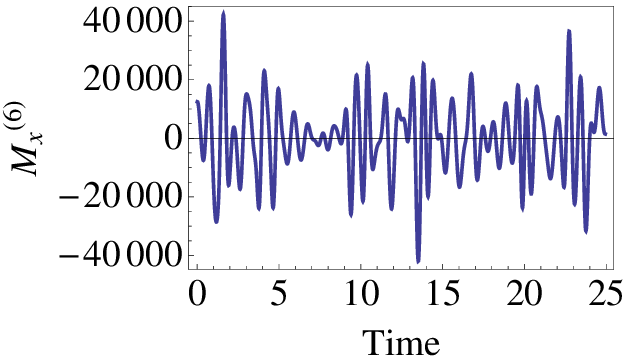,width=4.2cm } }
\put(143, 82.5){  \epsfig{file=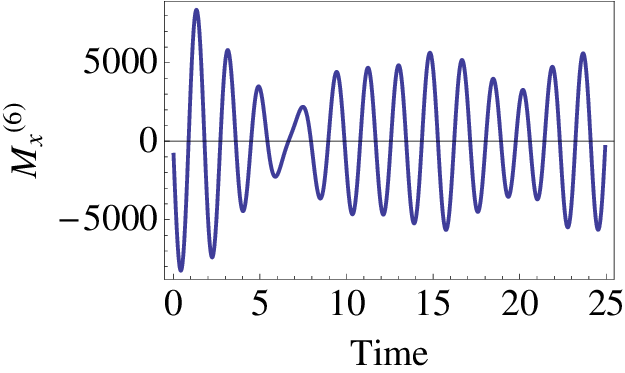,width=4.2cm } }

\put(100, 55.5){  \epsfig{file=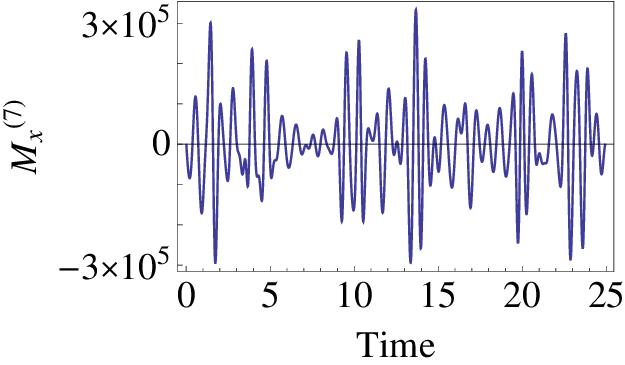,width=4.2cm } }
\put(143, 55.5){  \epsfig{file=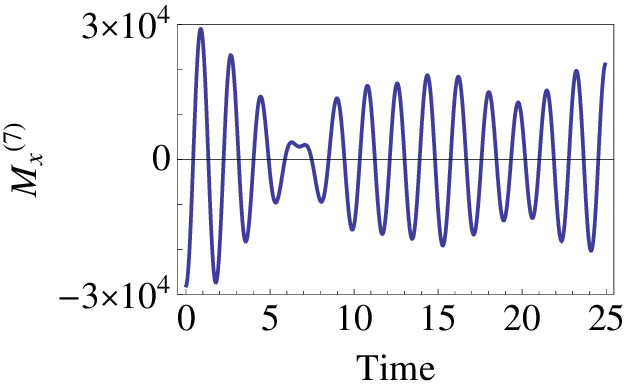,width=4.2cm } }

\put(100, 28.5){  \epsfig{file=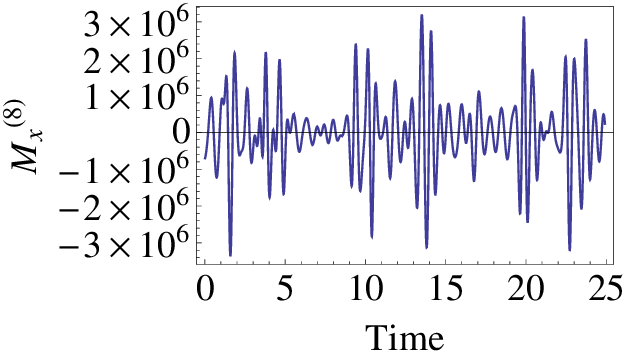,width=4.2cm } }
\put(143, 28.5){  \epsfig{file=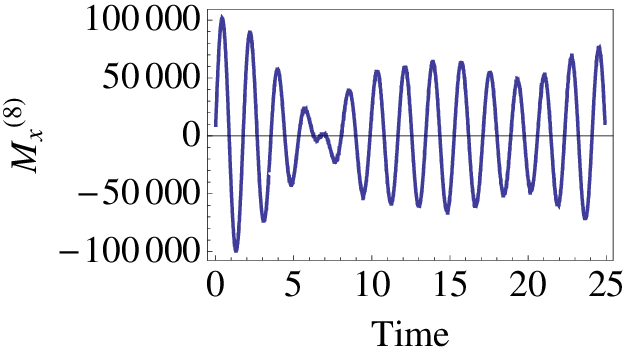,width=4.2cm } }

\put(100, 1.5){  \epsfig{file=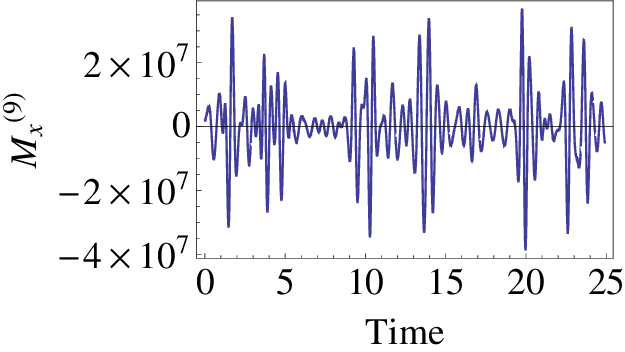,width=4.2cm } }
\put(143, 1.5){  \epsfig{file=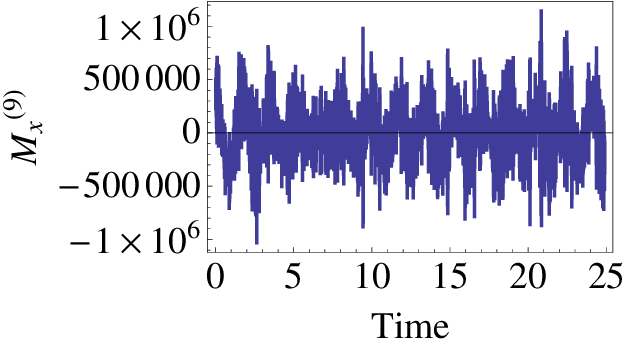,width=4.2cm } }

\put(119,138){Chaotic}
\put(162,138){Integrable}

}
\end{picture} 

\caption
{Plots of $M_x(t)$ and its nine derivatives for the chaotic system (left) and the integrable system (right). Note: the 9th derivative for the integrable case exhibits extrinsic noise due to numerical rounding errors. }
\label{derivatives}

\end{figure*} 

\end{widetext}

The numerical derivatives beginning with the ninth in the integrable case and tenth in the chaotic case show signs of numerical noise associated with the accumulated rounding errors. In order to exclude this concern,  the presentation in Fig.~\ref{main} was limited to the 7th derivative.

\section{Details on the calculation of the power spectra in Fig. \ref{main}}
\label{details}
 
The power spectra in Fig. \ref{main}  were obtained by calculating the square of the absolute value of the  Discrete Fourier Transform of the respective time series multiplied by a smooth-shaped window $\{ w_i \}$  to mitigate the spectral leakage from low frequencies to high frequencies due to the finite length of the time series. That is,
if the original time series is $ \left\{ ...,  \left( t_i, M_{x, i}^{(n)} \right),  ... \right\}$, then the modified time series is 
$ \left\{ ...,  \left( t_i, w_i M_{x, i}^{(n)} \right),  ... \right\}$.
We used the ten-percent Tukey window \cite{tukey} defined as:

 
\[
w_i=0.5\left\{ 1- \hbox{cos}\left[ \frac{2 \pi  i}{0.1N_t} \right] \right\}  \textup{ \ \ for \ \ $0 \leq i \leq 0.05N_t$ } , 
\]
\vspace{1mm}
\[
w_i =1 \textup{ \ \  for \ \ $0.05N_t \leq i \leq 0.95 N_t$},
\]
\vspace{1mm}
\[
w_i =0.5\left\{ 1- \hbox{cos}\left[ \frac{2\pi (i-N_t)}{0.1N_t} \right] \right\}  \textup{ \ \ for \ \ $0.95 N_t \leq i \leq N_t$}, 
\] 
where $N_t$ is the index of the last discretized time point in $\{t_0,...,t_i,...,t_{N_t} \}$. This window  smoothly suppresses the time series to zero at $t_0$ and $t_{N_t}$ to reduce the finite-length effects. 

We did not use the above procedure for the power series presented in Fig. \ref{short}, because, in that case, the time series was very short ($T=10$), and, as a result, the window function $w_i$ would contaminate the relevant part of the power spectrum.

\begin{figure} \setlength{\unitlength}{0.1cm}
\begin{picture}(88,57) 
{
\put(3, 0){  \epsfig{file=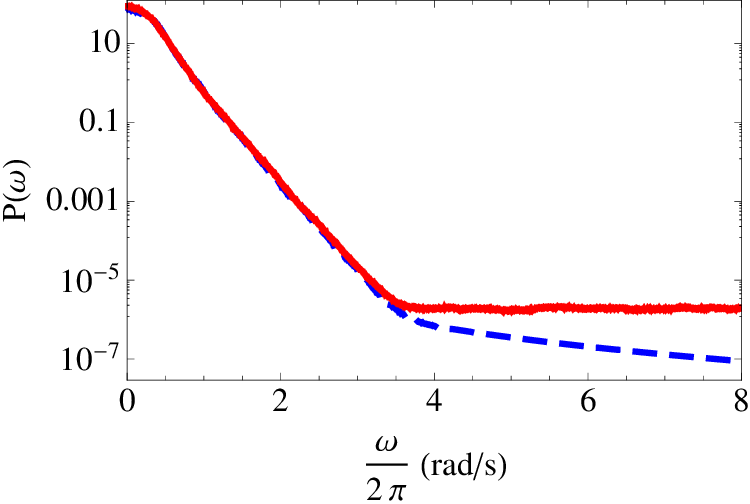,width=8cm } }
}
\end{picture} 

\caption{ \label{fig-precision}  (Color online) Comparison between two power spectra for the same setting as in Fig.~\ref{main}(e) obtained with single machine precision (solid red line) and double machine precision (dashed blue line).}
\end{figure}

We have also checked that the exponential tails of the power spectra are not the consequence of the rounding errors accumulated in the course of the numerical simulation routine. For this purpose, we computed two power spectra for the same system using either single or double machine precision.  As shown in Fig.~\ref{fig-precision}, the two spectra coincide in the range of frequencies not affected by the spectral leakage. 

\section{Truncated Toda lattice}
\label{truncated}

\begin{figure} \setlength{\unitlength}{0.1cm}

\begin{picture}(88 , 26 ) 
{
\put(0, 0){  \epsfig{file=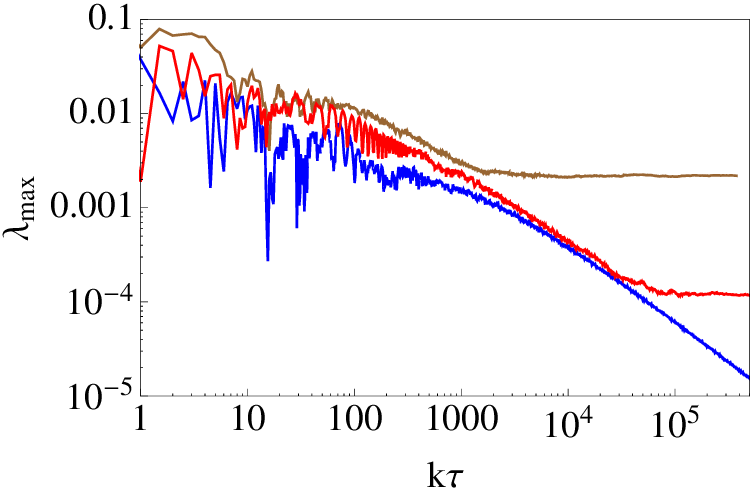,width=4.1cm } }
\put(44, 0){  \epsfig{file=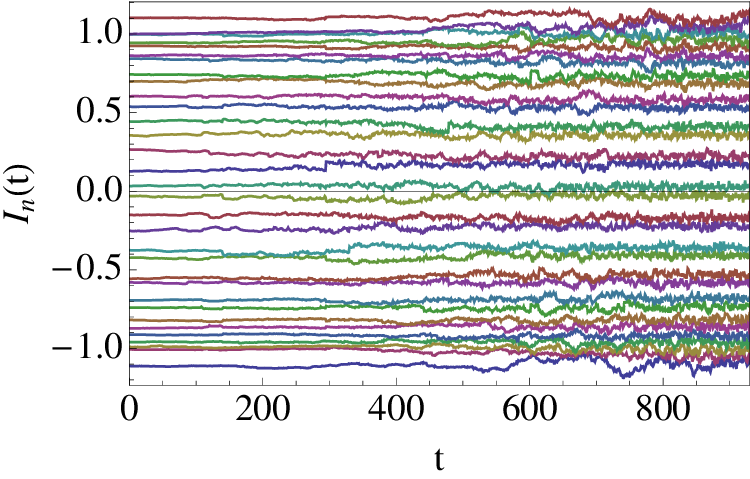,width=4.cm } }
\put(-0.5, 24){{ (a)}}
\put(42.5, 24){{ (b)}}
}
\end{picture} 

\caption{ \label{lamda-toda}  (Color online) (a)  $\lambda_{max}(k\tau)$ as defined in Eq. \ref{lambdamax}, computed for a Toda lattice consisting of 32 particles with exponential potential (blue),  exponential potential truncated up to the third order (red) and exponential potential truncated up to the fifth order (brown). (b) The eigenvalues of the $L$ matrix defined in Eq. \ref{L-matrix} for the third  order truncated potential.}
\end{figure}

A truncation of the exponential potential for the Toda lattice at any order higher than two leads to chaotic dynamics \cite{yoshida88}.  We illustrate this property numerically by computing $\lambda_{max}(k\tau)$ as defined in Eq. \ref{lambdamax}  for the exponential potential, third-order  truncated exponential potential, and fifth-order truncated exponential potential as shown in Fig. \ref{lamda-toda} (a). The initial conditions were selected randomly in all cases. We note that  $\lambda_{max}(k\tau)$ decays steadily as a power law in the first case, indicating a vanishing Lyapunov exponent, while it saturates at finite values of  $\lambda_{max}(k\tau)$ in the last two cases. In Fig. \ref{lamda-toda} (b) we depict the values of $I_n$ for the second case where we see evidently that the eigenvalues of $L$ are no longer constants of motion.

\end{document}